\documentclass[12pt]{article}
\usepackage{epsfig}

\usepackage{amssymb}
\usepackage{amsmath}
\usepackage{amsfonts}
\usepackage{amsthm}

\theoremstyle{theorem}
\newtheorem{lem}{Lemma}

\theoremstyle{definition}

\newtheorem{pozn}{Remark}

\def\bp{\begin{proof}}
\def\ep{\end{proof}}

\def\be{\begin{equation}}
\def\ee{\end{equation}}
\def\ba{\begin{array}{c}}
\def\ea{\end{array}}

\def\ben{$$}
\def\een{$$}

\newcommand{\bea}{\begin{eqnarray}}
\newcommand{\eea}{\end{eqnarray}}

\begin{document}

\titlepage

\vspace{.35cm}

 \begin{center}{\Large \bf

$N-$site-lattice analogues of  $V(x)={\rm i}x^3$

  }\end{center}

\vspace{10mm}

 \begin{center}

 {\bf Miloslav Znojil}

 \vspace{3mm}
Nuclear Physics Institute ASCR,

250 68 \v{R}e\v{z}, Czech Republic

{e-mail: znojil@ujf.cas.cz}

\vspace{3mm}

%\today, kinemjpa.tex

\end{center}

\vspace{5mm}

%\newpage

%{\it ad hoc} finite-lattice
%enables us to construct non-equivalent metrics in closed form.
%  and deformation

\section*{Abstract}

A discrete $N-$level alternative to the popular imaginary cubic
oscillator is proposed and studied. As usual, the unitarity of
evolution is guaranteed by the introduction of an {\it ad hoc},
Hamiltonian-dependent inner-product metric, the use of which defines
the physical Hilbert space and renders the Hamiltonian (with real
spectrum) observable. Due to the simplicity of our model of dynamics
the construction of the set of eligible metrics is shown tractable
by non-numerical means which combine the computer-assisted algebra
with the extrapolation and/or perturbation techniques.

\vspace{5mm}

\subsection*{KEYWORDS}

cryptohermitian discrete Schroedinger equations; deformed imaginary
cubic oscillators; exceptional points; spectral topology; critical
exponents; operators of metric; extrapolations; Fibonacci numbers;

\newpage
 \section{Introduction \label{zacatek} }

Many measured spectra of energies may be interpreted as excitations
of a quasiparticle. The simplest fits of such a type employ the
elementary one-dimensional differential Schr\"{o}dinger equation
 \be
 -\frac{d^2}{dx^2}\,\psi_n(x)+
  V(x)\,\psi(x) =E_n\,\psi_n(x)\,
 \ \ \ \ \ \ \
 \psi(\pm \Lambda)=0\,,\ \ \ \ \Lambda \leq \infty
 \label{SEloc}
 \ee
containing a  real potential $V(x)$. In more sophisticated models,
potential $V(x)$ may even be allowed complex, provided only that the
spectrum itself remains real (cf. reviews
\cite{Dorey,Carl,ali,SIGMA} for details).

Whenever $V(x) \neq V^*(x)$, the reality of the spectrum may be
fragile and sensitive to perturbations  \cite{fragile}. A remarkable
exception emerges with the robustly real spectra generated by many
potentials with the property $V(x) = V^*(-x)$ called, in the
literature, ${\cal PT}-$symmetry \cite{Carl} alias
parity--pseudo-Hermiticity \cite{ali} alias Krein-space-Hermiticity
\cite{Tretter}.

In the early studies of this remarkable mathematical phenomenon the
attention of the authors has mainly been restricted to the imaginary
cubic oscillator example $V(x)=V^{(IC)}(x)={\rm i}x^3$ modified,
possibly, by some other, asymptotically subdominant terms (cf.,
e.g., papers by Caliceti et al \cite{Caliceti}, by Alvarez
\cite{Gabriel} or by Bender et al \cite{BM,BB} and several further
authors \cite{BMM}). In what follows we shall mainly feel inspired
by this choice of $V(x)$ as well.

The essence of our present message will lie in the recommendation of
a drastic simplification of the necessary mathematics. In a way
explained in section \ref{expla} this will be achieved by means of
the replacement of the differential Schr\"{o}dinger equation
(\ref{SEloc}) by its discrete, difference-equation analogue. A few
simplest illustrations will be then added in section~\ref{energies}
where the discretization of the coordinate will be shown to
facilitate the study of the reality (i.e., in principle,
observability) of the spectrum.

The desirable flexibility of our present discrete simulation of the
dynamical energy spectra will be shown achieved by a supplementary
one-parametric deformation of the potential resembling slightly the
influential proposal of deformation ${\rm i}x^3 \to ({\rm
i}x)^{3+\delta}$ by Bender and Boettcher \cite{BB}. The details will
be described in section \ref{deformation}. The subsequent section
\ref{metrics} will then recall the known theory and explain some of
its details via the simplest possible example with $N=2$. In
sections \ref{nje4} -- \ref{nje8} we shall finally present some
applications of this theory to the models with $N=4$, $N = 6$ and
general $N=2K \geq 8$, respectively.  Section \ref{summary} is
summary.

\section{Discrete Schr\"{o}dinger equations \label{expla}}

The spectrum of many Krein-space-Hermitian Hamiltonians $H\neq
H^\dagger$ has been found robustly real and bounded below
\cite{BB,DDT,Geza}. In the spirit of the general theory as outlined,
first, by Scholtz et al \cite{Geyer}, one can conclude that the
apparent non-Hermiticity is ``false" and that it may be
reinterpreted as a mere consequence of an inappropriate choice of
the Hilbert-space representation $L^2(\mathbb{R})\equiv {\cal
H}^{(F)}$. Hence, the abstract remedy is straightforward and lies in
an interpretation-mediating transition to a ``standard" Hilbert
space of states ${\cal H}^{(S)}$ \cite{SIGMA}.

In the practical applications of such a theoretical scheme the
original vector space is usually being endowed with a general, {\em
non-local} inner product,
 \be
 \langle \psi|\phi \rangle^{(S)}=
 \int\int\,\psi^*(x)\, \Theta(x,y)\,\phi(y)\,dx\,dy \equiv
  \langle \psi|\Theta|\phi \rangle^{(F)}\,, \ \ \ \
  \Theta=\Theta^\dagger>0\,.
  \label{innpro}
 \ee
For the one-dimensional imaginary cubic oscillator Hamiltonians
$$H=H^{(IC)}(\lambda)=-\nabla^2+{\rm i}\lambda x^3+\ldots$$ (where the
dots may represent certain asymptotically subdominant terms) the
first constructions of the necessary ``standard metric"  were
perturbative. They appeared in 2003 \cite{firstcubic} and in 2004
\cite{Bcubic}, yielding the metric operator
 \be
 \Theta=\Theta(H^{(IC)})=\exp(-Q_1\lambda-Q_2
 \lambda^2-\ldots)
 \,
 \label{brody}
 \ee
with $Q_1(x,y)={\rm i} xy(x^2+y^2){\rm sign} (x-y)/2^{4}$, etc. In
the light of the well known fact that the assignment of the metric
$\Theta$ to a given Hamiltonian $H$ cannot be unique in general
\cite{Geyer,SIGMAdva}, additional, multiparametric metrics
$\Theta_{{(c_1,c_2,\ldots)}}(H^{(IC)})$ were further found and
constructed in Refs. \cite{Bcubic} and \cite{cubic}.

The latter calculations proved facilitated by an additional
assumption of the absence of a nontrivial fundamental length $L>0$
in the theory. Unfortunately, the {\em presence} of such a
preassigned length scale has been found essential in
Ref.~\cite{scatt} where an extension of the formalism to the
scattering dynamical regime has been proposed. Naturally, under the
assumption of the presence of a fixed length scale $L>0$, additional
ambiguities will emerge in the metrics. In the differential-operator
models their explicit specification might prove prohibitively
difficult (cf., e.g., Ref.~\cite{Jonesdva} or section 6 of
Ref.~\cite{cubic} for related comments).

We shall address here this problem while accepting the
discretization strategy of Ref.~\cite{scatt}. In the way based on
the use of equidistant, Runge-Kutta grid-point coordinates
 $
  x_k=-\Lambda+k\,h\,, \ k = 0,  1, \ldots,
 N+1$, the
ordinary differential Schroedinger Eq.~(\ref{SEloc}) will be
replaced by its discrete version
 \be
 -\frac{\psi(x_{k-1})-2\,\psi(x_k)+\psi(x_{k+1})}{h^2}+V(x_k)\,
 \psi(x_k)
 =E\,\psi(x_k)\,
 \label{SEdis}
 \ee
with $
  x_{N+1}=\Lambda$ (i.e.,  $h=2\Lambda/(N+1)$) and with the Dirichlet
   boundary conditions
  \be
 \psi(x_{0})=\psi(x_{N+1})=0\,.
 \label{bcbs}
 \ee
This leads to the reduction of the problem to the (numerical)
determination of the $N-$plets of eigenvalues $E^{(N)}_j$, $j =
1,2,\ldots,N$, i.e., of the eigenvalues
$\varepsilon_j:=h^2E^{(N)}_j\in (0,4)$ of the re-scaled,
$N-$dimensional Hamiltonians
 \be
 {H}^{(N)}=
 \left (
 \begin{array}{ccccc}
 2+h^2V(x_{1})&-1&&&\\
 -1&2+h^2V(x_{2})&-1&&\\
  &-1& 2+h^2V(x_{3}) & \ddots&\\
  & &\ddots&\ddots&-1 \\
 &&&-1&2+h^2V(x_{N})
 \ea
 \right )\,.
 \label{dvanact}
 \ee
We intend to make use of this definition of the Hamiltonian in the
whole rest of our present paper. Nevertheless, before we fully
concentrate on its bound-state aspects and consequences, let us make
a small detour and point out that our present acceptance of the
Dirichlet boundary conditions (\ref{bcbs}) is in fact just one of
the two basic alternative options which are at our disposal in the
phenomenologically oriented applications. In a very close parallel
to the continuous cases, these boundary conditions could have been
replaced, even in the Runge-Kutta-discretized models, by the
alternative scattering scenario.

In the latter context, one might feel discouraged by a threatening
technical difficulty connected with the necessity of working with
the infinite-dimensional matrices. Still, there exist tricks which
enable us to avoid these difficulties by assuming that our
interaction $V(x)$ in Eq.~(\ref{dvanact}) is just short-ranged
(plus, admissibly, non-Hermitian and non-local). For more details
and/or for an explicit constructive illustration of such an
alternative possibility, interested readers might consult, e.g.,
Ref.~\cite{Jonestri}.

\section{The series of one-parametric Hamiltonians\label{energies}}

Let us pick up the potential $V(x_j)={\rm i}x_j^3$ and insert it in
Eq.~(\ref{dvanact}). Once we abbreviate $a=h^5/8$ we obtain the
following one-parametric sequence of Hamiltonians
 \ben
 H^{(2)}(a)=
 \left (
 \begin{array}{cc}
 2-{\rm i}a&-1\\
 -1&2+{\rm i}a
 \ea
 \right )\,,\ \ \
 H^{(3)}(a)=
 \left (
 \begin{array}{ccc}
 2-8{\rm i}a&-1&0\\
 -1&2&-1\\
 0&-1&2+8{\rm i}a
 \ea
 \right )\,,
 \een
 \be
 H^{(4)}(a)=
 \left (
 \begin{array}{cccc}
 2-27{\rm i}a&-1&0&0\\
 -1&
 2-{\rm i}a&-1&0\\
 0&-1&
 2+{\rm i}a&-1\\
 0&0&-1&2+27{\rm i}a
 \ea
 \right )\,,
  \ldots\,.
 \label{dva}
 \ee
The first element $H^{(2)}(a)$ of this series is particularly
elementary. For this reason it has also been chosen and studied in
the methodical part of Ref.~\cite{firstcubic} (cf. its section II.
B).

For the sake of brevity we shall restrict our attention  to the
matrices with even dimension $N=2K$. Thus, in the simplest case with
$K=1$ one encounters the compact and explicit energy formula
$\varepsilon_{1,2}=2\mp \sqrt {1-{a}^{2}}$. Such a two-level
spectrum  remains real iff $a \in (-1,1)$, forming a circle in the
energy-coupling plane. The similar formulae and conclusions may be
also obtained at the next few $K$s.

%\subsection{Numerical subtleties: $N\leq 6$ }

%\newpage
%********** Figure 1 zde
\begin{figure}[h]                     %instead of \begin{figure}[t]
\begin{center}                         %instead of \begin{center}
\epsfig{file=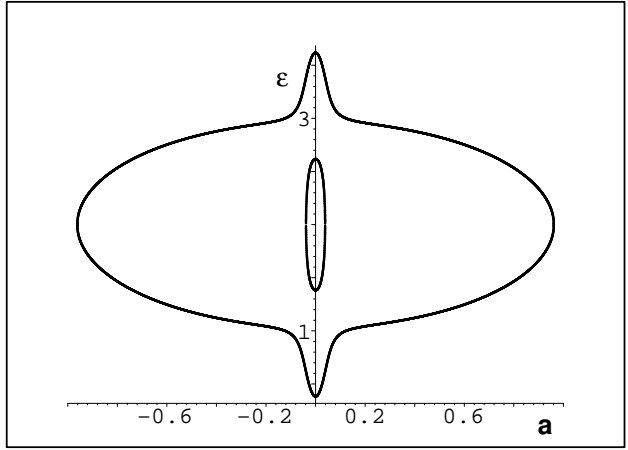,angle=270,width=0.6\textwidth}
\end{center}                         %instead of \end{center}
\vspace{-2mm} \caption{The parameter-dependence of the {\em
real-energy} eigenvalues $\varepsilon(a)$ of the toy Hamiltonian
$H^{(4)}(a)$ [= the third item in the list (\ref{dva})]. In
topological language, this spectral locus is formed of the two
(deformed) cocentric circles.
 \label{firm4}}
\end{figure}
%%\newpage

 %\noindent
The first nontrivial sample of the $a-$dependence of the spectrum
may be obtained at $N=4$. Its shape is displayed in
Fig.~\ref{firm4}. The algebraic representation of these energies is
elementary,
 $$
 \varepsilon_{1,2,3,4}=
2\mp 1/2\,\sqrt {6 -1460\,{a}^{ 2}\pm 2\,\sqrt
{529984\,{a}^{4}-1680\,{a}^{2}+5}}\,.
 $$
One can easily deduce that these energies remain real for $a \in
\left (-\alpha^{(4)},\alpha^{(4)}\right )$ where
$\alpha^{(4)}=1/2-1/18\,\sqrt {69}\approx  0.0385208965$.

%\subsection{$N=6$}

%\newpage
%********** Figure 1 zde
\begin{figure}[h]                     %instead of \begin{figure}[t]
\begin{center}                         %instead of \begin{center}
\epsfig{file=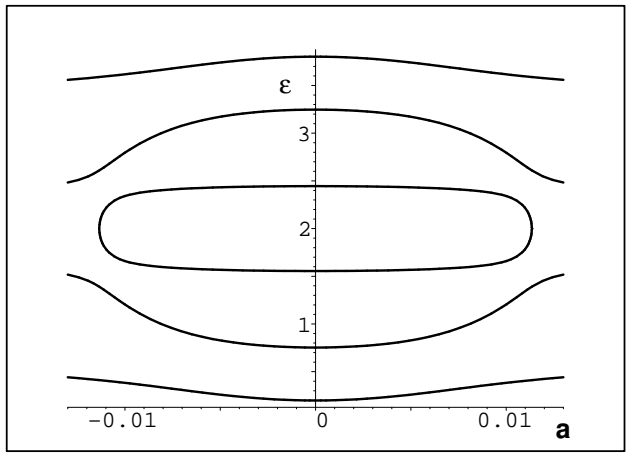,angle=270,width=0.6\textwidth}
\end{center}                         %instead of \end{center}
\vspace{-2mm} \caption{Same as Figure \ref{firm4}, at the next even
dimension $N=6$. The spectral locus is formed of the three
(deformed) cocentric circles (note that the horizontal axis is
rescaled).
 \label{firm6}}
\end{figure}
%%\newpage

 %\noindent
Numerical difficulties start emerging at $N=6$ since the purely
algebraic representation of the spectrum requires the use of Cardano
formulae in which the imaginary numerical errors occur and survive,
disappearing only in the infinite-precision arithmetics. Still, the
more or less routine control of precision enables us to conclude
that the $N=6$ spectrum remains real iff $a \in \left
(-\alpha^{(6)},\alpha^{(6)}\right )$. The approximate numerical
value of $\alpha^{(6)} \approx 0.011344897$ may be read out of
Fig.~\ref{firm6} and/or of its appropriate systematic
magnifications.

%%\newpage

%%\newpage

%\newpage
%********** Figure 1 zde
\begin{figure}[h]                     %instead of \begin{figure}[t]
\begin{center}                         %instead of \begin{center}
\epsfig{file=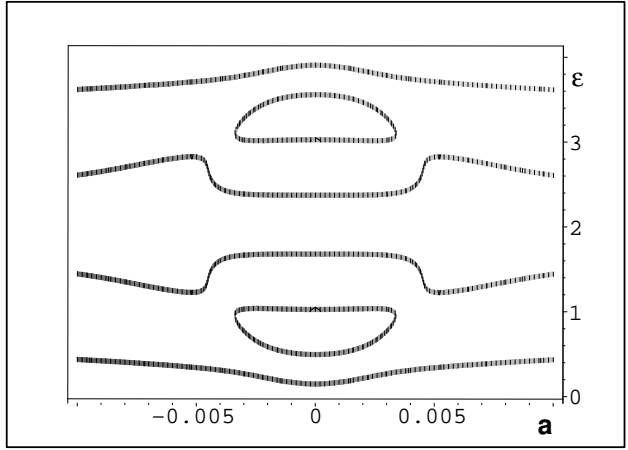,angle=270,width=0.6\textwidth}
\end{center}                         %instead of \end{center}
\vspace{-2mm} \caption{The next item in the series of Figures
\ref{firm4} and \ref{firm6}. The horizontal axis is further rescaled
and the $N=8$ spectral locus gets nontrivial. In topological terms,
it becomes composed of the vertical array of the three circles, all
being circumscribed by the fourth one.
 \label{firm8}}
\end{figure}

%\newpage
%********** Figure 1 zde
\begin{figure}[h]                     %instead of \begin{figure}[t]
\begin{center}                         %instead of \begin{center}
\epsfig{file=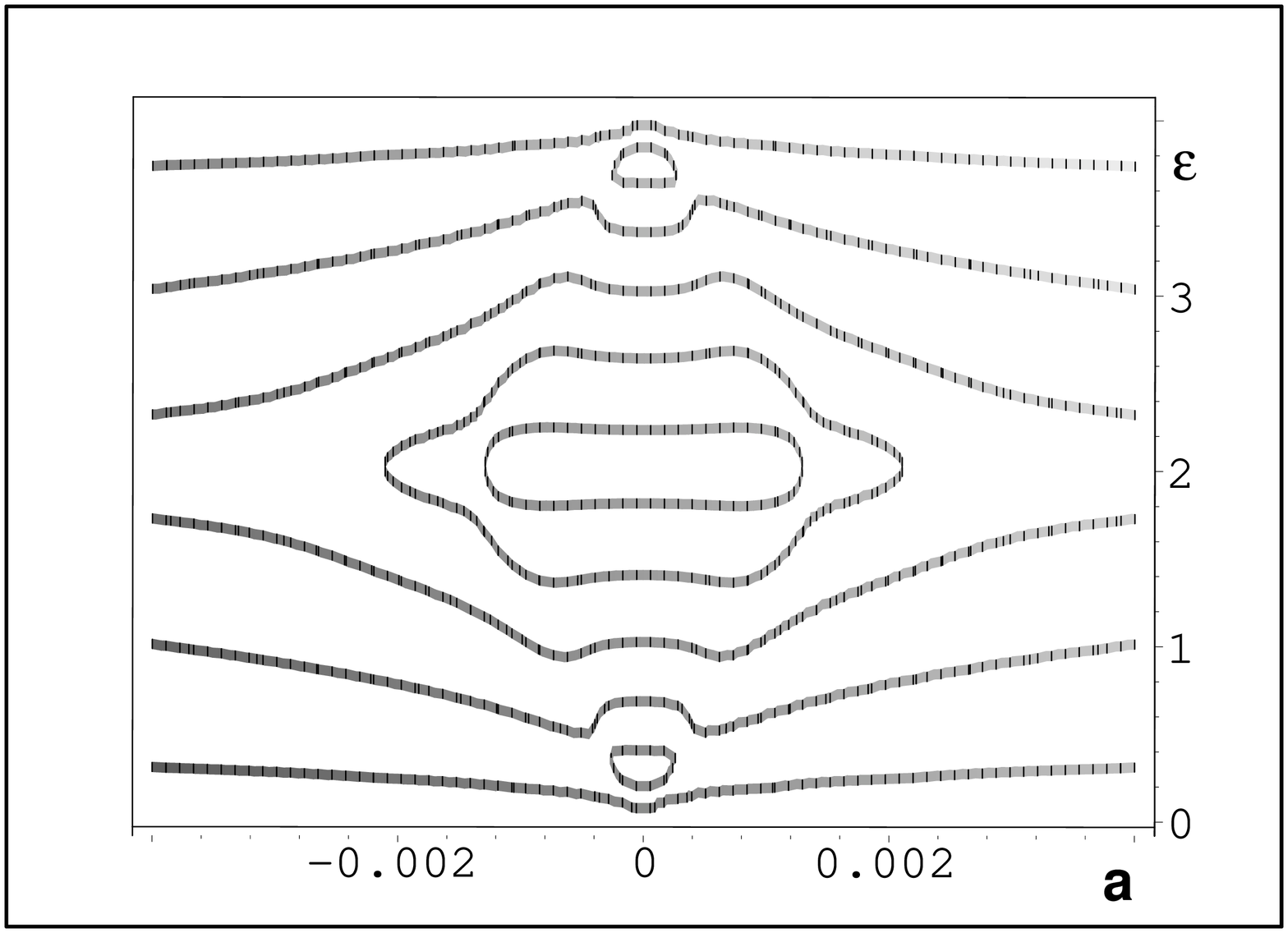,angle=270,width=0.6\textwidth}
\end{center}                         %instead of \end{center}
\vspace{-2mm} \caption{A sample of a ``multilevel descendant" of the
preceding  Figures. At $N=14$ we see a ``generic" pattern in which
we encounter $(N-6)/2$ concentric circles plus a single ``top" and
single ``bottom" circle, all being circumscribed by the ``biggest"
last circle. The interval of the reality of {\em all} of the
eigenvalues
 $\varepsilon(a)$ of the toy Hamiltonian
$H^{(14)}(a)$ has further shrunk.
 \label{firm14}}
\end{figure}
%%\newpage

 %\noindent

We may notice that at $N=4$ and $N=6$ the ends $\pm \alpha^{(4,6)}$
of the interval of the reality of the whole spectrum are determined
by the confluence (followed by the complexification) of a single
pair of energies (lying in the middle of the spectrum). Both of
these models are exceptional. Starting from $N=8$ the pattern gets
changed and we encounter the less trivial $a-$dependence of the
energies $\varepsilon(a)$ characterized by the fragility and
confluence of the {\em two} pairs of the energy levels (cf.
Fig.\ref{firm8}).

The latter form of the loss of reality is generic. The $N=14$ sample
of the spectrum as displayed in Fig.~\ref{firm14} elucidates some
details. In the picture we see that the spectral locus (i.e., the
set of all of the real energies $\varepsilon_j(a)$ with $a \in
(-\infty,\infty)$ and $j = 1, 2, \ldots, N$) preserves the form
which remains up-down and left-right symmetric. During the growth of
$N=2K$ the graphical analysis of the model remains feasible and
reveals a steady shrinking of the interval of the parameters $a$ for
which the spectrum remains real.

%\subsection{The intervals of reality of the spectra}

\begin{table}[t]
\caption{The exceptional-point coordinates of the loss of reality of
the first and second excited state ($z=3$). } \label{pexp4}

\vspace{2mm}

\centering
\begin{tabular}{||c||c|c||}
\hline \hline
    {\rm dimension} & critical &
     {\rm twice-degenerate}
    \\
    $N$ &  parameter $\alpha$&
       energy $\varepsilon_2(\alpha)=\varepsilon_3(\alpha)$
    \\
 \hline \hline
 2&  1  & 2\\
 4&  $1/2-\sqrt {69}/18$ & 2 \\
 6&  0.011344897 & 2 \\
 8&  0.003383828  & 0.9285\\
 10  &  0.001246890& 0.6194\\
 12&  0.000543788 & 0.4438 \\
 14&  0.000266880 & 0.3335 \\
 \vdots&
 \vdots&
 \vdots\\
 \hline
 \hline
\end{tabular}
\end{table}

A quantitative account of the latter phenomenon is presented in
Table~\ref{pexp4}. The inspection of this Table reveals that the
presence of the fundamental-length parameter $\alpha^{(N)}$ (= a
maximum of admissible $a$s, decreasing with $N$) in our present
model implies an obvious mismatch between the large$-N$ spectrum
(with a very small interval of reality) and the robustly real
spectrum which has been proved to exist in the differential-equation
limit  $N = \infty$ \cite{DDT}.

An explanation of the paradox is twofold. Firstly, the enormously
small magnitude of $a={\cal O}(h^5)$ makes the comparison only
sensible in the limit $a\to 0$ (in this sense, the spectrum of the
discrete model remains robustly real at all $N$). Secondly, at any
finite dimension $N< \infty$ even the role of the Runge-Kutta error
terms ${\cal O}({h^4})\gg {\cal O}({a})$ remains unspecified.
Whenever the limit of $N\to \infty$ is considered, this observation
offers another argument against drawing any $N \to \infty$
implications from the observations made at the finite $N$ and
nonvanishing values of $a$ in our models.

%Deformations $x^3$ $\to$ odd $f(x)$
\section{The series of two-parametric
Hamiltonians\label{deformation}}

In the examples of the preceding section we encounter another
suspicious feature even at any fixed and finite dimension $N = 2K$.
Indeed, the critical left and right points $a = \pm \alpha^{(2K)}$
of the loss of the reality of the spectrum seem to be exclusively
connected with the confluence of the first and second lowest excited
states $\varepsilon_{2}$ and $\varepsilon_{3}$ or, symmetrically, of
their equally ``privileged" mirror partners $\varepsilon_{N-2}$ and
$\varepsilon_{N-1}$.

In what follows we shall explain this apparent privilege as an
artifact caused by the too specific choice of the potential. The
constructive explanation will be based on a  rescaling of the
interaction $V(x)={\rm i}x^3$. Its third power term will be replaced
by an odd function  $f_{(z)}(x)\sim {\rm sign}(x) \,x^z$ with a real
exponent $z\in \mathbb{R}$. As long as our models are discrete, we
need not follow the conventional wisdom and require that the
functions $f_{(z)}(x)$ are analytic. Thus, we arrive at the family
of two-parametric Hamiltonians $ H^{(2K)}(a,z)=$
 \be
  \left[ \begin {array}{cccccccc}
 2-i\,a\,(2K-1)^z&-1&0&\ldots&&&\ldots&0
 \\
 -1&\ddots&\ddots&\ddots&&&&\vdots
 \\
 0&\ddots &2-i\,a\,3^z&-1&0&&&
 \\
 \noalign{\medskip}
 \vdots &\ddots&-1&2-ia&-1&0&&
 \\
 \noalign{\medskip}
 &&0&-1&2+ia&-1&\ddots&\vdots
 \\
 \noalign{\medskip}
 &&&0&-1&2+i\,a\,3^z&\ddots &0
 \\
 \vdots &&&&\ddots&\ddots&\ddots&-1
 \\
 \noalign{\medskip}
 0&\ldots&&&\ldots&0&-1&2+i\,a\,(2K-1)^z
 \end {array}
 \right]\,.
 \label{zzz}
 \ee
We shall see below that such a class of discrete deformations
$$V(x)={\rm i}x^3:={\rm i}f_{(3)}(x)\ \ \to \ \ V(x)={\rm i}f_{(z)}(x)$$
may be made responsible for the emergence of a full range of
possible complexification patterns. In the language of dynamics and
phenomenology, we shall show below that the emergence of the new
parameter $z$ in models (\ref{zzz}) renders them tractable as
offering a certain {\em complete} set of alternative, tuneable
topological features of the spectrum. In this sense our present {\em
complex-interaction} models may be perceived as complementing the
recent attempts of achieving a topology-related tuning of spectra
via a {\em non-locality} of real $V(x)$ \cite{graphs} or,
alternatively, via a {\em nontriviality} of the ``coordinates" $x$
living, say, on real closed loops \cite{graphsce} or on certain
specific multisheeted complex curves \cite{graphsb}.

%\section{The changes of
%topology at the critical exponents $z$\label{nje6}}

\subsection{Spectral loci $\varepsilon^{(N,z)}(a)$: the $N=4$
prototype\label{unje6}}

As we indicated, the key source of interest in our two-parametric
models $H^{(2K)}(a,z)$ should be seen in the enhancement of the
flexibility of the $a-$dependence of the spectrum. Moreover, one can
expect that the choice of $z$ might re-assign the role of the most
fragile levels all along the spectrum.

In an expected verification of these hypotheses, let us now select
$N=4$ and turn attention to the two-parametric Hamiltonian
 \be
 H^{(4)}(a,z)=
 \left[ \begin {array}{cccc}
 2-i\,a\,3^z&-1&0&0\\\noalign{\medskip}-1&2-ia&-1&0
 \\\noalign{\medskip}0&-1&2+ia&-1
 \\ \noalign{\medskip}0&0&-1&2+i\,a\,3^z\end {array}
 \right]\,.
 \label{zz}
 \ee
Due to the simplicity of such a generalization of the purely cubic
discrete model (\ref{dva}) it may be easily shown, graphically, that
the spectral pattern displayed in Fig.~\ref{firm4} does not change
too much when the exponent-parameter $z$ starts to be different from
three. Indeed, with the steady growth of $z>3$ one merely reveals
that the small internal ellipse in the picture of Fig.~\ref{firm4}
will shrink. In a numerical test performed at $z=25$, for example,
we still recognized the existence of this inner ellipse but only on
the scale of $a$s of the order of magnitude of $10^{-12}$.

%\newpage
%********** Figure 1 zde
\begin{figure}[h]                     %instead of \begin{figure}[t]
\begin{center}                         %instead of \begin{center}
\epsfig{file=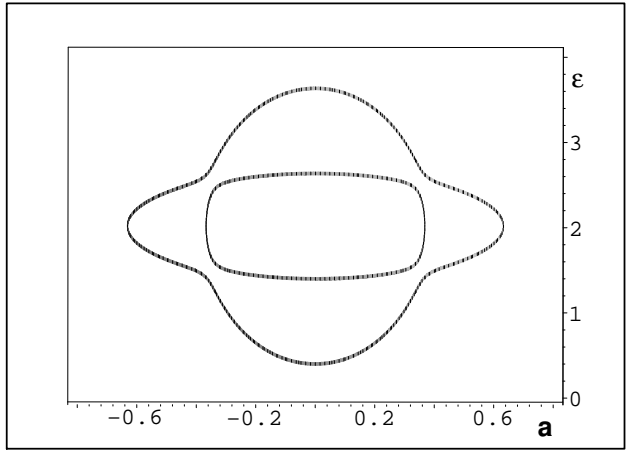,angle=270,width=0.6\textwidth}
\end{center}                         %instead of \end{center}
\vspace{-2mm} \caption{The $a-$dependence of the real eigenvalues
$\varepsilon(a)=\varepsilon^{(N,z)}(a)$ of the two-parametric toy
Hamiltonian $H^{(N)}(a,z)$ at $N=4$ [cf. Eq.~(\ref{zz})]. Slightly
above a critical value of  $z=85/64\sim 1.328125>z_{critical}$, the
pattern is still formed by the two cocentric circles.
 \label{firmzb}}
\end{figure}
%%\newpage

In the opposite direction of the change, i.e., with the decrease of
$z$ below three the internal ellipse broadens. From the inspection
of Fig.~\ref{firmzb} we may deduce that slightly below the value of
$z=85/64$ the internal ellipse must ultimately touch the external
deformed circle. At this moment the whole spectral locus will
acquire the form of the two intersecting ellipses. After the further
small decrease of $z$ we recognize a qualitatively (i.e.,
topologically) new pattern, the form of which is sampled, at the
next rational numerical value of $z=84/64$, in Fig.~\ref{firmzs}.

%\newpage
%********** Figure 1 zde
\begin{figure}[h]                     %instead of \begin{figure}[t]
\begin{center}                         %instead of \begin{center}
\epsfig{file=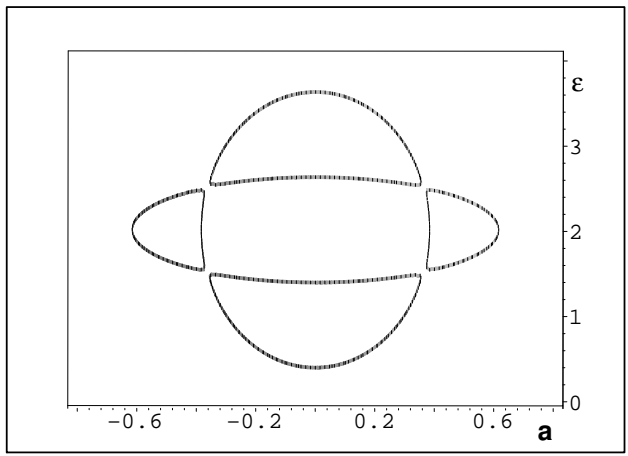,angle=270,width=0.6\textwidth}
\end{center}                         %instead of \end{center}
\vspace{-2mm} \caption{The breakdown of the cocentric-circles
pattern of Figure~\ref{firmzb} when the value of $z$ dropped just
slightly below the critical value,  $z=84/64=1.3125 \lessapprox
z_{critical}$.
 \label{firmzs}}
\end{figure}
%%\newpage

With the continuing decrease of $z$ we witness a quick shrinking of
the two spurious large-$|a|$ intervals of partial reality of the
spectrum. Both of them are centered around $|a|= 1/2$ while leaving
still the two levels real (cf. Fig.~\ref{firmzd} where we choose
$z=81/64=1.265625$). We also repeated the same graphical analysis
below the latter value of the exponent. We revealed that quickly,
both the anomalous partial-reality intervals have got empty and
disappeared.

%\newpage
%********** Figure 1 zde
\begin{figure}[h]                     %instead of \begin{figure}[t]
\begin{center}                         %instead of \begin{center}
\epsfig{file=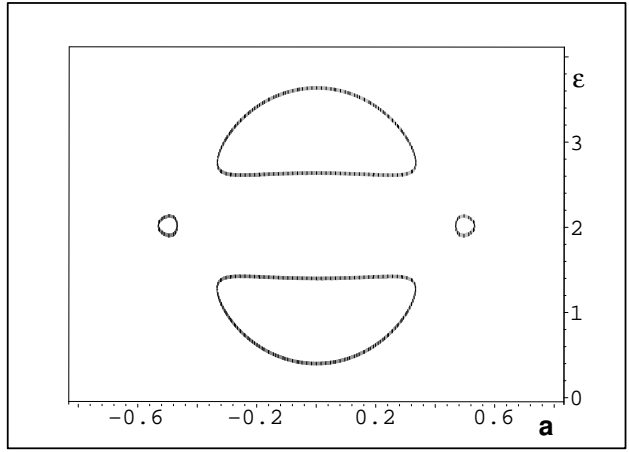,angle=270,width=0.6\textwidth}
\end{center}                         %instead of \end{center}
\vspace{-2mm} \caption{The ``evolution-to-disappearance" of the left
and right ``anomalous" topological circles of Figure~\ref{firmzs} at
$z=81/64=1.265625<z_{critical}$. During the further decrease of $z$,
 just the real spectral locus composed of the
horizontal array of two nonintersecting (deformed) circles will
survive.
 \label{firmzd}}
\end{figure}
%%\newpage

Needless to add that the further steady decrease of $z$  already
keeps the ``two-oval" topology of the spectral pattern unchanged. We
checked this empirical rule in the exactly tractable limit of $z=0$.
In this limit the quadruplet of the energies acquires the compact
form
 $$
 \varepsilon(a)= 2\pm 1/2\,\sqrt {6-4\,{a}^{2}\pm 2\,\sqrt
 {-16\,{a}^{2}+5}}\,,
 \ \ \ \ \ \ z=0\,.
 $$
One can strictly deduce that this spectrum remains real iff $a \in
(-\sqrt{5}/4,\sqrt{5}/4)$.

%\subsection{Discrete non-cubic $N=8$ power-law oscillators}
 %at $z=z_{k-th\
%critical}^{(N)}$

\subsection{Sequences of rearrangements: the $N=8$
prototype\label{nje6be}}

%\newpage
%********** Figure 1 zde
\begin{figure}[h]                     %instead of \begin{figure}[t]
\begin{center}                         %instead of \begin{center}
\epsfig{file=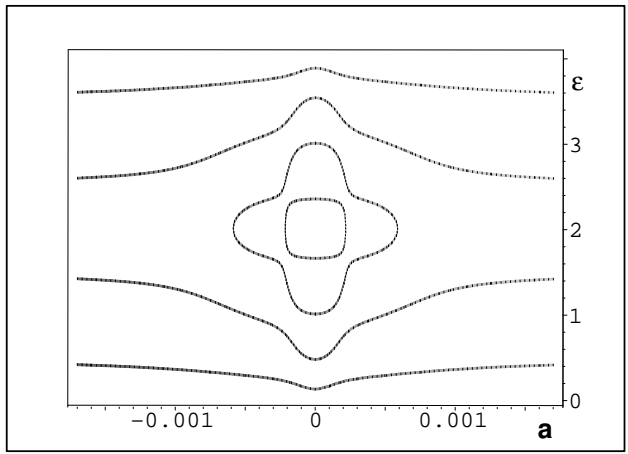,angle=270,width=0.6\textwidth}
\end{center}                         %instead of \end{center}
\vspace{-2mm} \caption{For the less elementary matrix
 $ H^{(2K)}(a,z)$ of Eq.~(\ref{zzz})
 with
$K=4$, the spectral locus remains composed of the four concentric
(deformed) circles at all the sufficiently large exponents $z$. This
pattern is sampled here at
 $z=9/2>z_{first\ critical}^{(8)}$.
 \label{odpo}}
\end{figure}
%%\newpage

 \noindent
The straightforward generalization of Eq.~(\ref{zz}) to any
dimension yields the $N$ by $N$ matrix $ H^{(2K)}(a,z)$  of
Eq.~(\ref{zzz}). The related geometric shapes of the spectra just
generalize the $N=4$ pattern. At any $N=2K$ and at all of the
sufficiently large exponents we revealed that the picture of the
real spectral locus in the $(\varepsilon,a)-$plane remains
topologically equivalent to the set of cocentric circles. At $N=8$
and $z=9/2$ and $z=8/2$, the respective Figs.~\ref{odpo} and
\ref{odcpo} offer the two characteristic samples of the evolution of
the large-exponent spectrum with the decrease of the exponent
$z>z_{first\ critical}^{(2K)}$.

%\newpage
%********** Figure 1 zde
\begin{figure}[h]                     %instead of \begin{figure}[t]
\begin{center}                         %instead of \begin{center}
\epsfig{file=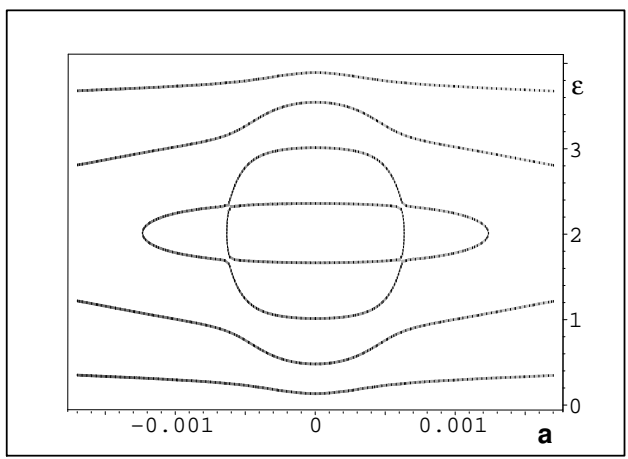,angle=270,width=0.6\textwidth}
\end{center}                         %instead of \end{center}
\vspace{-2mm} \caption{The topology of Figure \ref{odpo} is not yet
changed  in the ``close-to-crossing" case of $z=8/2 \gtrapprox
z_{first\ critical}^{(8)}$.
 \label{odcpo}}
\end{figure}
%%\newpage

In a series of graphical experiments using the smaller and smaller
rational exponents $z$ we were able to keep the numerical precision
under good control. We discovered that $z_{first\ critical}^{(8)}
\lesssim 4$. Below this value though safely above $z_{second\
critical}^{(8)} \gtrsim 12/4$ (cf. Fig.~\ref{firm8} above), one
encounters the new topological pattern sampled by Figs.~\ref{odcupo}
or~\ref{odcupodva}.

The corresponding second-pattern regime may further be split in two
stages. In the initial stage of the decrease of $z$ there survive
two large$-|a|$ anomalies which quickly shrink and, not too far
below the first critical value of the exponent, disappear. In the
final stage of the decrease the picture of the whole real spectral
locus in the $(\varepsilon,a)-$plane acquires the ``fully canonical"
form of the two cocentric  (deformed) circles with a vertically
ordered pair of non-intersecting circles (or rather deformed
ellipses) inside.

%\newpage
%********** Figure 1 zde
\begin{figure}[h]                     %instead of \begin{figure}[t]
\begin{center}                         %instead of \begin{center}
\epsfig{file=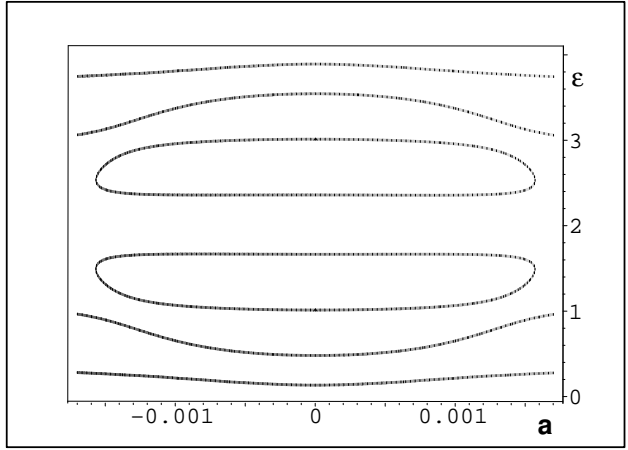,angle=270,width=0.6\textwidth}
\end{center}                         %instead of \end{center}
\vspace{-2mm} \caption{The innermost circles of Figures \ref{odpo}
or \ref{odcpo} are being replaced by their horizontal array  at
$z=14/4\in (z_{second\ critical}^{(8)},z_{first\ critical}^{(8)})$.
Notice that we moved safely below $z_{first\ critical}^{(8)}$ so
that the temporary left and right real-eigenvalue anomalies have
already (and very quickly) disappeared.
 \label{odcupo}}
\end{figure}
%%\newpage

%\newpage
%********** Figure 1 zde
\begin{figure}[h]                     %instead of \begin{figure}[t]
\begin{center}                         %instead of \begin{center}
\epsfig{file=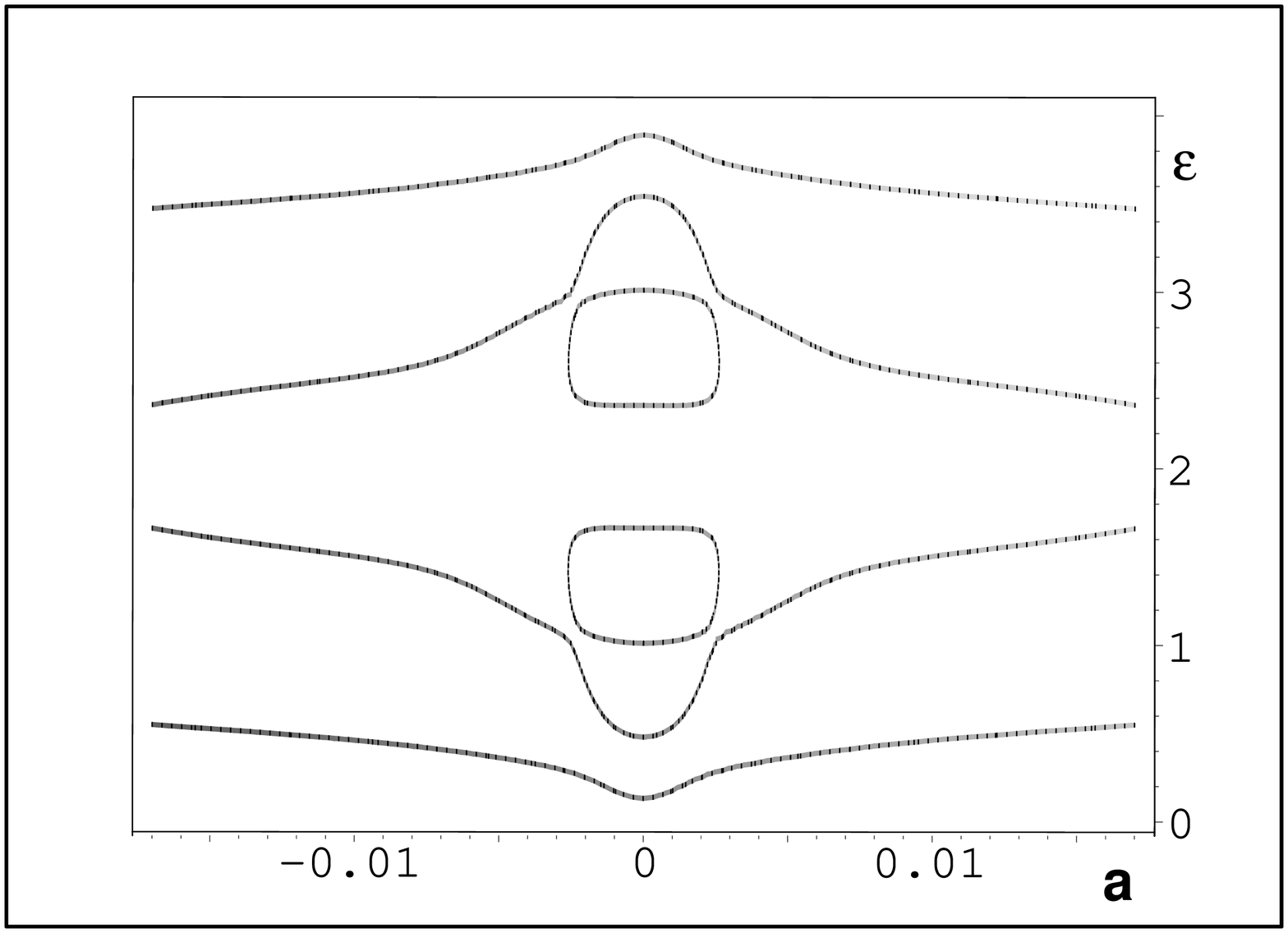,angle=270,width=0.6\textwidth}
\end{center}                         %instead of \end{center}
\vspace{-2mm} \caption{The topology-preserving deformation of
Figure~\ref{odcupo} during the further decrease of the exponent to
 $z=13/4\gtrsim z_{second\
critical}^{(8)}$ (notice also the change of scale of $a$).
 \label{odcupodva}}
\end{figure}
%%\newpage

The next, second change of the topological pattern has been spotted
to occur between $z=13/4$  (cf.~Fig.~\ref{odcupodva}) and $z=7/4$
(cf.~Fig.~\ref{upw}) while the third change certainly follows
between $z=7/4$ and $z=6/4$ (cf.~Fig.~\ref{downw}). Ultimately, the
list of the topological changes of the $N=8$ spectral loci is made
complete during the transition between $z=6/4$ and  $z=1/2$
(cf.~Fig.~\ref{firm8jp}). One arrives at the other, small-exponent
extreme in which the spectral locus stays unchanged and
topologically equivalent to a vertically ordered quadruplet of
separate deformed circles.

%\newpage
%********** Figure 1 zde
\begin{figure}[h]                     %instead of \begin{figure}[t]
\begin{center}                         %instead of \begin{center}
\epsfig{file=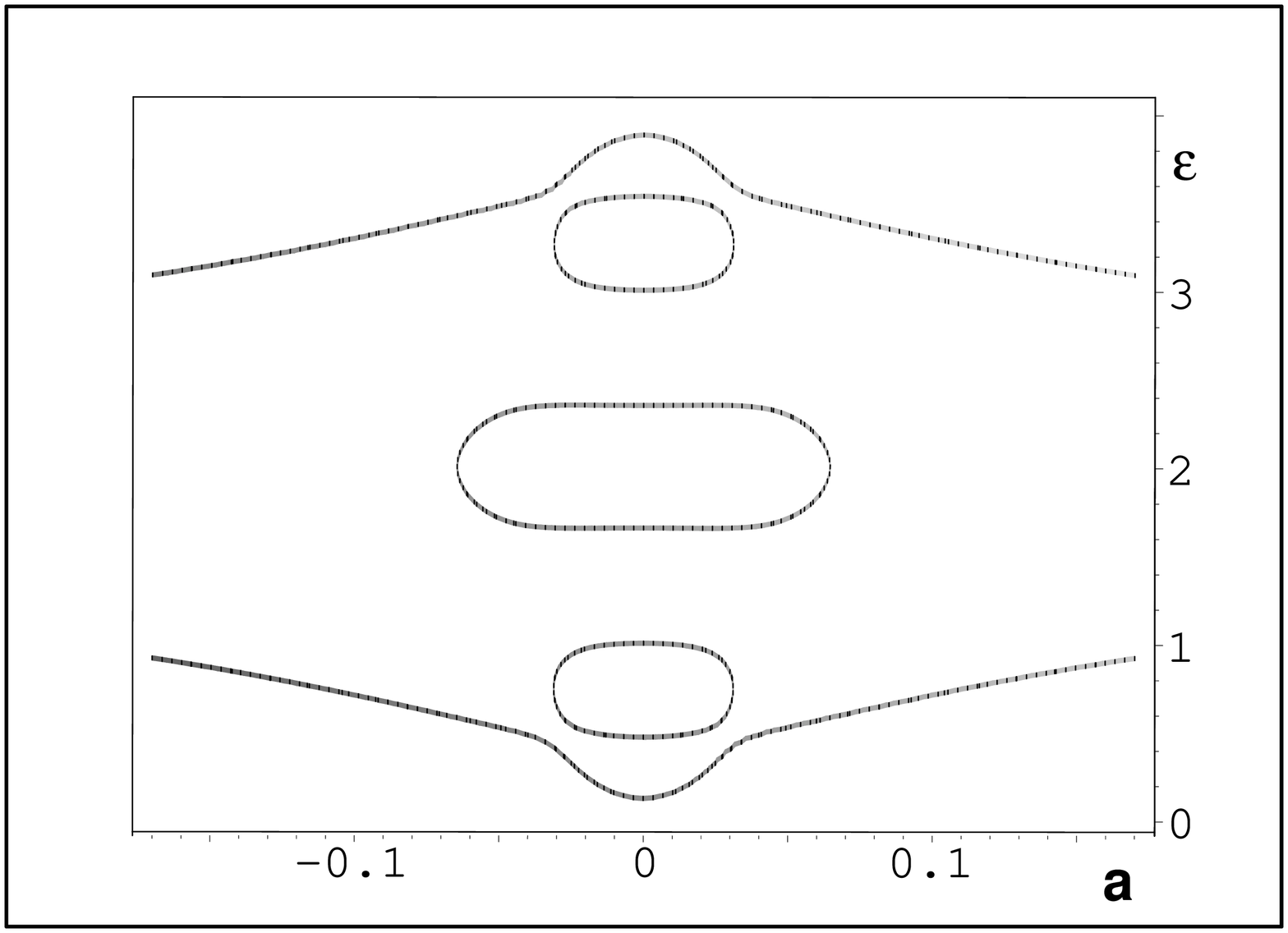,angle=270,width=0.6\textwidth}
\end{center}                         %instead of \end{center}
\vspace{-2mm} \caption{The next change of topology of the spectral
locus as sampled at $z=7/4> z_{third\ critical}^{(8)}$.
 \label{upw}}
\end{figure}
%%\newpage

%\newpage
%********** Figure 1 zde
\begin{figure}[h]                     %instead of \begin{figure}[t]
\begin{center}                         %instead of \begin{center}
\epsfig{file=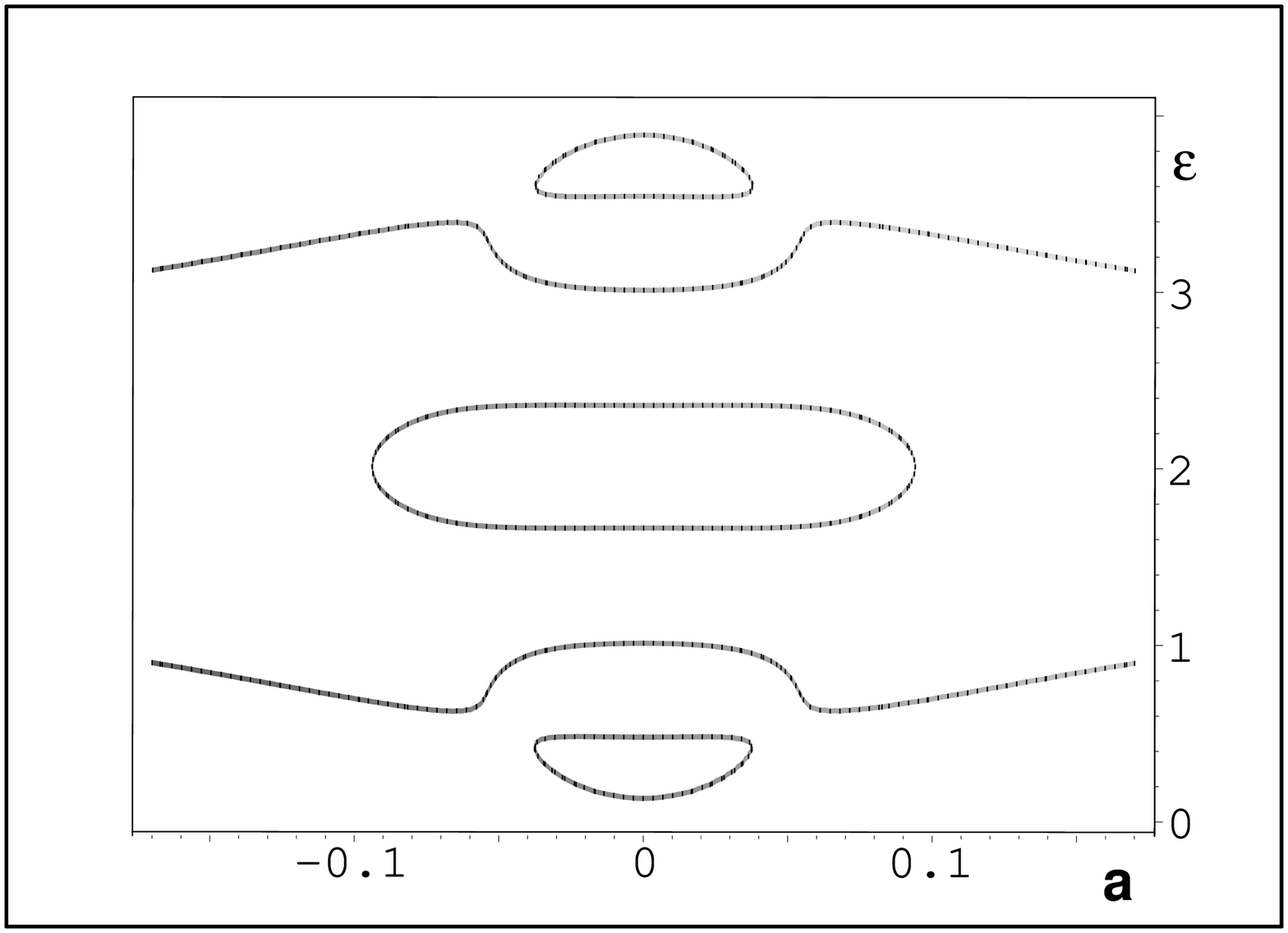,angle=270,width=0.6\textwidth}
\end{center}                         %instead of \end{center}
\vspace{-2mm} \caption{The next topological arrangement of the
spectral locus sampled at $z=6/4 \in ( z_{third\
critical}^{(8)},z_{fourth\ critical}^{(8)})$.
 \label{downw}}
\end{figure}
%%\newpage

The results of our calculations aimed at a more precise
determination of the numerical values of the quadruplet of critical
exponents at $N=8$ are summarized in Table~\ref{roxp4}. In order to
make these calculations feasible, we were forced to give up the full
control of the numerical precision. For this reason,
Table~\ref{roxp4} does not contain any estimates of the error bars
so that even the last digits in our numerical values of the critical
exponents $z$ should not be taken for granted.

Marginally, let us add that, strictly speaking, our present list of
the five critical points might have been complemented not only by
the list of certain ``secondary critical" points $\tilde{z}$
(marking the disappearance of the above-mentioned left and right
partial-reality intervals in the anomalous, large$-|a|$ dynamical
regime) but also by the list of certain ``tertiary critical" points
$\tilde{\tilde{z}}$ at which these left and right anomalies change
from ``containing" to ``not-containing" an even narrower
four-level-reality subinterval.

The readers who would be interested in the similar subtleties might
consult the results of this type as obtained, e.g., in
Ref.~\cite{graphs}. In the present context, a sample of such an
analysis has merely been performed in the small-exponent regime with
$z\geq 0$ where we found  $z_{fourth\ critical}^{(8)}\approx
1.0358$.

In this setting the unchanged topology without anomalies has been
only demonstrated to exist in the slightly smaller interval of $z\in
(-\infty,\tilde{z})$ with $\tilde{z}\approx 1.033$. Within the
latter, anomaly-free  interval, the variations of $z$ still
preserved the strict reality just along the four non-intersecting
ellipses, more and more deformed and all the time located in a
vertical arrangement. In the adjacent, very short interval of $z\in
(1.033,1.0358)$ there emerged the left and right anomaly as sampled,
at $N=4$, in Fig.~\ref{firmzd}.

At the fourth critical value of $z\approx 1.0358$ we obtained, in
accord with our expectations, the transitional intersection of
ellipses as sampled, at $N=4$, in Fig.~\ref{firmzs}. We can conclude
that our numerical study of the $N=8$ model confirmed that when we
ignore the partial-reality anomalies as inessential for the fully
unitary quantum systems, the range of the exponents $z\geq 0$ (or
rather $z>-\infty$) splits into five subintervals on which the
spectral loci $\varepsilon(a)$ become topologically non-equivalent.
The exhaustive and fully explicit description of these $N=8$
topologies is also given in Table~\ref{roxp4}.

%\newpage
%********** Figure 1 zde
\begin{figure}[h]                     %instead of \begin{figure}[t]
\begin{center}                         %instead of \begin{center}
\epsfig{file=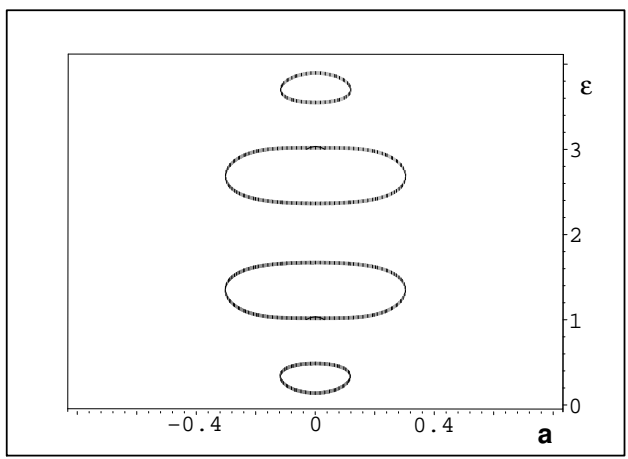,angle=270,width=0.6\textwidth}
\end{center}                         %instead of \end{center}
\vspace{-2mm} \caption{The ultimate, ``small-exponent" topology
which is sampled here at $z=1/2$ but which is exhibited by any
spectral locus $\varepsilon(a)$ of matrix (\ref{zzz}) at $K = 4$ and
any real $z<z_{fourth\ critical}^{(8)}$.
 \label{firm8jp}}
\end{figure}
%%\newpage

\begin{table}[t]
\caption{The five intervals of
exponents %\\
%$z\in (0,\infty)=
%(0,c_1)\bigcup(c_1,c_2)\bigcup(c_2,c_3)\bigcup(c_3,c_4)
%\bigcup(c_4,\infty)$ \\
yielding topologically non-equivalent real spectral loci
$\varepsilon(a)$ at $N=8$.} \label{roxp4}

\vspace{2mm}

\centering
\begin{tabular}{||l|l|l||}
\hline \hline
     {\rm the graph of  $\varepsilon_j=\varepsilon_j(a),\,\ j=1,2,\ldots,8$}
     &{\rm illustration}&
   {\rm interval}\\
   \hline
   \hline
      {\rm four cocentric  (deformed) circles}&Figs.~\ref{odpo},~\ref{odcpo}&
    $z\in (3.982,\infty)$ \\
      {\rm two in vertical array, encircled by two}&Figs.~\ref{odcupo},~\ref{odcupodva}
      &
    $z\in (3.178,3.982)$ \\
      {\rm three in vertical array, encircled by one}&Fig.~\ref{upw}&
    $z\in (1.630,3.178)$ \\
      {\rm vertical, but the  middle pair cocentric}&Fig.~\ref{downw}&
    $z\in (1.0358, 1.630)$ \\
               {\rm four (deformed) circles in vertical array}
               &Fig.~\ref{firm8jp}&$z\in (-\infty,1.0358)$ \\
 \hline
 \hline
\end{tabular}
\end{table}

%\subsection{Discrete non-cubic $N=8$ power-law oscillators}
 %at $z=z_{k-th\
%critical}^{(N)}$

\subsection{Rearrangements: Fibonacci-sequence connection\label{nje6ce}}

Fig.~\ref{odpo} may be perceived as the inessential $N=8$
sophistication of the $N=4$ pattern of Fig.~\ref{firmzb}. Similarly,
Fig.~\ref{odcpo} recycles the structure which is shown in
Fig.~\ref{firmzs}, while Fig.~\ref{odcupo} may be read as
paralleling Fig.~\ref{firmzd}. The generic $N > 6$ pattern only
starts emerging, at $N=8$, during the transition from
Fig.~\ref{odcupodva} to Fig.~\ref{upw}. It is characterized by the
{\em two} parallel confluences of energy pairs at the critical
exponent $z_{second\ critical}$. The similar phenomenon occurs at
all of the further critical points, viz., at the remaining two
topology-changing transitions from  Fig.~\ref{upw} to
Fig.~\ref{downw} and from Fig.~\ref{downw} to Fig.~\ref{firm8jp}.

At any higher dimension $N=2K$ the similar sequence of changes of
topology can be assigned a fully systematic description. Thus, once
we choose any $N$ and assume the complete knowledge of the set
${\cal T}_N$ of all of the related spectral-locus arrangements
(consisting of the $K-$plets of deformed circles) which are
topologically non-equivalent, we may contemplate a constructive
transition to the next dimension $N'=2K+2$. In the first step we
build the first subset ${\cal F}_{N'}$ of new schemes ${\cal
T}_{N'}$ as the set of elements of the old set ${\cal T}_N$ which
are merely encircled by an additional, single outer (deformed)
circle.

The construction of ${\cal T}_{N'}= {\cal F}_{N'}\bigoplus {\cal
S}_{N'}$ may be then completed by the discovery of the one-to-one
correspondence of the second part ${\cal S}_{N'}$ of the new set of
schemes to the ``one step older" set ${\cal T}_{N''}$ where
$N''=2K-2$. Each of the ``older" elements only has to be
complemented by the {\em pair} of the single upper and the single
lower additional (deformed) circles, indeed.

In the enumeration of the complete sets of non-equivalent patterns
by induction, it is now sufficient to verify that the numbers
$\#{\cal T}_{2K}$ of the elements of the sets ${\cal T}_{2K}$ are
equal to $K$ for $K=1,2,3$. Thus, we have proved

\begin{lem}
We have $\#{\cal T}_{2K}=F_{K-1}$ where  $F_j$ is the $j-$th
Fibonacci number.
\end{lem}

\begin{pozn}
The sequence of Fibonacci numbers $F_j$ is defined by recurrences
$F_j=F_{j-1} + F_{j-2}$ yielding $\# {\cal T}_{8}=5$, $\#{\cal
T}_{10}=8$, $\#{\cal T}_{12}=13$, etc.
\end{pozn}

%\newpage

\section{Interpretation\label{metrics}}

Usually \cite{Ghosh} people decide to work in a fixed, specific
Hilbert space ${\cal H}^{(S)}$ and treat a given quantum system as
physical if and only if its time evolution is unitary. Naturally,
this is the strategy which does not change if the space ${\cal
H}^{(S)}$ proves endowed with a general, nontrivial metric $
\Theta=\Theta^\dagger>0$ (i.e., with its inner product defined in
terms of this metric, cf. Eq.~(\ref{innpro}) above). One still has
to guarantee that every candidate for an operator of observable
proves self-adjoint {\em with respect to this metric} \cite{Geyer}.

For our present, manifestly non-Hermitian matrix representations
$H^{(N)}$ of the Hamiltonians (which will have to play the role of
the generators of the {\em unitary} time evolution in  ${\cal
H}^{(S)}$) we must guarantee their Hermiticity with respect to the
given nontrivial metric. Such a form of nontrivial Hermiticity could
be better called cryptohermiticity \cite{SIGMA}. In the context of
mathematics, such a condition is often being interpreted as the
Dieudonn\'{e}'s \cite{Dieudonne} ``quasi-Hermiticity" constraint,
 \be
   H^{(N)}=\left
 [H^{(N)}\right ]^\ddagger:=
 \Theta^{-1}\left [H^{(N)}\right ]^\dagger\Theta\,.
 \label{dieudonne}
 \ee
In the application of such an approach which is to be employed in
what follows, we shall always start from a given matrix $H^{(N)}$
and reconstruct the {\em ad hoc} metric (or metrics)
$\Theta=\Theta(H)$ via Eq.~(\ref{dieudonne}), treating this
constraint as a linear algebraic set of equations for the matrix
elements of the metric.

\subsection{The simplest $N=2$ illustration}

For the most elementary, $z-$independent $N=2$ ``input" Hamiltonian
(cf. the first item in Eq.~(\ref{dva})), the solution of
Dieudonn\'{e}'s constraint (\ref{dieudonne}) yields the complete,
two-parametric family of the most general matrices of ``output"
pseudometrics,
 \be
 \Theta\left [H^{(N)}(a) \right ]=\Theta\left [H^{(N)}(a) \right
 ]_{(k,m)}=\left[ \begin {array}{cc} k&km-ika
 \\\noalign{\medskip}km+ika&k
 \end{array} \right]\,,\ \ k,m \in \mathbb{R}\,.
 \label{pse2}
 \ee
Physical condition $a \in (-1,1)$ guaranteeing the reality of the
energies must be complemented by the condition of positivity of the
(two) eigenvalues
 \be
 \theta=\theta_{\pm}=k\pm \sqrt {{k^2m}^{2}+{k}^{2}{a}^{2}}>0\,.
 \label{positivitym}
 \ee
Only such a condition will open the possibility of using the
corresponding pseudometric (\ref{pse2}) as a metric in ${\cal
H}^{(S)}$ \cite{Geyer}.

We see that $k$ must be positive  and larger than the square root.
We may set $a=\cos \beta\,\sin \gamma$ and $m=\cos \beta\,\cos
\gamma$ with, say, $\beta\in (0,\pi)$ and $\gamma\in (0,\pi)$ and
$-1<\cos \beta<1$. This reparametrization leads to the final and
most general positive definite and Hermitian metric
 \be
 \Theta=
 \Theta\left \{H^{(2)}[a(\beta,\gamma)]
 \right \}_{[k,m(\beta,\gamma)]}=k\cdot\left[
 \begin {array}{cc} 1&e^{-{\rm i}\gamma}\cos \beta
 \\\noalign{\medskip}e^{{\rm i}\gamma}\cos \beta&1
 \end{array} \right]\,
 \label{metrika2}
 \ee
which is never diagonal. As long as it contains free parameters,
their choice will fix the inner product and, hence, it will specify
the Hilbert space ${\cal H}_{(\beta,\gamma)}^{(S)}$ of admissible
states of the model. The choice of parameters $\beta$ and $\gamma$
will determine {\em both} the Hamiltonian $H^{(2)}(a)$ and the
metric $\Theta_{(k,m)}^{(S)}$ (with, say, $k=1$), i.e., the true and
complete {\em physical} contents of the theory.

%select, say, $\beta\in (0, \pi)$ and $\gamma\in (-\pi/2, \pi/2)$ and

\subsection{Eligible observables}

In our $N=2$ theory, {\em any} matrix
$\Lambda=\Lambda^{(2)}_{(\beta,\gamma)}$ representing an observable
quantity must be self-adjoint in ${\cal H}_{(\beta,\gamma)}^{(S)}$,
i.e. \cite{SIGMA}, we must have
 \be
 \Lambda_{(\beta,\gamma)}^\dagger\,\Theta_{[k,m(\beta,\gamma)]}
 =\Theta_{[k,m(\beta,\gamma)]}\,\Lambda_{(\beta,\gamma)}\,.
 \label{diela}
 \ee
This equation must be satisfied by the eligible representation
matrices
 \be
 \Lambda=\left[
 \begin {array}{cc}
 G+{\rm i}g&B+{\rm i}b
 \\\noalign{\medskip}C+{\rm i}c&D+{\rm i}d
 \end{array} \right]\,.
 \label{leila}
 \ee
Naturally, Eq.~(\ref{diela}) admits an arbitrary $k-$rescaling of
$\Theta\left \{H^{(2)}[a(\beta,\gamma)]\right \}$ and/or a similar
trivial rescaling of $\Lambda_{(\beta,\gamma)}$. In terms of matrix
elements one can easily check that this equation imposes four real
constraints upon the eight free parameters in (\ref{leila}). This
means that the $N=2$ family of available observables is
four-parametric in general.

The details of the construction are left to the readers. We can only
summarize that the three constraints are trivial and that they
merely define quantities $B$, $C$ and the difference $G-D$. The
remaining, fourth constraint acquires the form of a linear relation
between sums $c_\Sigma=b+c$ and $g_\Sigma=g+d$ with the unique
solution  $g_\Sigma=0$. One can conclude that for a given input
$m=m(\beta,\gamma)$ and $a=a(\beta,\gamma)$ the final and entirely
general form of the $N=2$ observable reads
 \be
  \Lambda=\Lambda(D,b,c,g)=\frac{1}{a}\cdot \left[ \begin {array}{cc}
   D\,a- b-c+i\,g\,{a}\,,
   &
   g-b\, m+i \,b\,a
 \\\noalign{\medskip}
 {g+c\,m}+i\,c\,a\,,
 &
 D\,a-i\,g\,a
 \end {array} \right]\,.
 \label{geleila}
 \ee
We can check that our original Hamiltonian is reobtained at $D=2$
(i.e., $G=2$), $b=c=0$  (i.e., $B=C=-1$)  and $g= -a\ ( =-d)$.

In Ref.~\cite{Geyer} the authors recommended to proceed in an
opposite direction and select a few matrices (or operators)
$\Lambda_1, \Lambda_2, \ldots$ of observables in advance (say, on
some fitting or  phenomenological grounds). Naturally, these input
matrices must necessarily be self-adjoint in ${\cal H}^{(S)}$, i.e.,
in our $N=2$ example, proportional to our general formula
(\ref{geleila}). Thus, in the generic case, the series of the
necessary cryptohermiticity conditions will, sooner or later,
specify all of the values of the free parameters in the metric (up
to the above-mentioned trivial rescaling of course).

\subsection{The concept of charge}

In the so called ${\cal PT}-$symmetric quantum mechanics \cite{Carl}
one introduces an additional requirement which may be mathematically
interpreted as the assumption of the Hermiticity of the Hamiltonian
in a suitable Krein space \cite{Tretter,Langer}. Formally speaking,
one just preselects an indefinite Krein-space metric ${\cal P}$ with
the property ${\cal P}^2=I$ (called, conventionally, ``parity"). One
then postulates the following indefinite-metric parallel of
Eq.~(\ref{dieudonne}),
 \be
 \left [H\right ]^\dagger{\cal P}={\cal P}\,H
 \,.
 \label{pseudodieudonne}
 \ee
The main benefit of such an auxiliary assumption is seen in the
possibility of the construction of a special and {\em unique} metric
$\Theta^{({\cal CPT})}={\cal CP}$ called ${\cal CPT}$ metric. The
new operator ${\cal C}$ with property ${\cal C}^2=I$ is being
interpreted as a charge~\cite{BBJ}.

Our present toy-model Hamiltonian matrices $H=H^{(N)}(a,z)$ satisfy
Eq.~(\ref{pseudodieudonne}) with the Hamiltonian-independent
matrices of parity ${\cal P}={\cal P}^{(N)}$ containing just the
unit elements along the secondary diagonal (i.e., ${\cal
P}^{(N)}_{m,n}= 1$ iff $m+n=N+1$ while ${\cal P}^{(N)}_{m,n}= 0$
otherwise). Thus, at $N=2$ one could treat ${\cal P}^{(2)}$ as an
indefinite limit of $\Theta\left [H^{(2)}(a) \right ]$ with
$k^{({\cal P})}  \to 0$ while $m^{({\cal P})} \to \infty$ and
$k^{({\cal P})}m^{({\cal P})} \to 1$. From the factorization
requirement $\Theta^{({\cal CPT})}={\cal CP}$ we may eliminate the
complex charge
 \be
 {\cal C}=\Theta\left [H^{(2)}(a) \right ]_{(k,m)}{\cal P}^{(2)}=\left[
 \begin {array}{cc}
 u&v
 \\\noalign{\medskip}y&z
 \end{array} \right]\,
 \label{uleila}
 \ee
and get $v=y=k$ and $z=u^*=ke^{i\gamma}\cos \beta$ (cf.
Eq.~(\ref{metrika2})). The $N=2$ version of condition ${\cal C}^2=I$
requires not only that  $\gamma=\gamma^{({\cal CPT})}= \pi/2$ (i.e.,
that $a=\cos \beta$) but also that $\beta=\beta^{({\cal CPT})}$ is
such that $\sin \beta^{({\cal CPT})}=1/k$. The resulting metric is
unique and we have the unique charge
 \be
 {\cal C}^{({\cal CPT})}=k \cdot \left[
 \begin {array}{cc}
 - {\rm i} a& 1
 \\\noalign{\medskip}1& {\rm i} a
 \end{array} \right]\,.
 \label{uleila}
 \ee
The comparison of this formula with the general specification of an
observable (\ref{geleila}) reveals that the charge is an observable
in which $D=b=c=0$ and $g = - \cos \beta/\sin \beta=- \sqrt{k^2-1}$.

\section{The first nontrivial $N=4$ model\label{nje4}}

\subsection{The complete set of pseudometrics }

The general four-parametric Hermitian $N=4$ candidates for the
metric (= ``pseudometrics") may be obtained, most easily, from
Eq.~(\ref{dieudonne}) again. These solutions appear symmetric with
respect to their second diagonal. They may be written in the closed
four-parametric form of matrix $\Theta\left [H^{(4)}(a,z) \right
]_{(k,m,r,h)}=$
 \be
 =\left[ \begin {array}{cccc}
 k&m-ikw&W^*&Z^*
  \\\noalign{\medskip}m+ikw&r&h-i
 \left( kw+ra \right)&W^*
 \\\noalign{\medskip}W &h+i
 \left( kw+ra \right) &r&m-ikw
 \\\noalign{\medskip} Z&W &m+ikw&k
 \end {array} \right]
 \label{gene4}
 \ee
where we introduced function $w=w(z,a) = {3}^{z}{a}\ $  (this
quantity must be positive and, for $z> 0$, larger than $a$) and
where we abbreviated
 $$
 W=W(k,m,r)=-{w}^{2}k+r-k- kwa+i \left( wm+ma \right)\,,
 $$
 $$
 Z=Z(k,m,r,h)=m{a}^{2}-{w}^{2}m-m+h-i \left( kw-ka-kw{a}^{2}-rw+ {w}^{3}k
 \right)\,.$$
This result seems to indicate that the above-noticed full-matrix
structure of the $N=2$ metrics will survive the transition to any
dimension. Our present choice of the discrete but strictly {\em
local} non-Hermitian interactions $V(x_j)$ only seems to admit the
non-band-matrix, strongly nonlocal forms of {\em all} of the
Hermitizing metrics.

\subsection{The condition of positivity \label{n4a}}

Although an immediate correspondence between the above-displayed
general $N=4$ pseudometric with the most common and usual metric
$\Theta=I$ seems out of question, we may still select, for
illustration purposes, the unit main diagonal in formula
(\ref{gene4}), $k=r=1$. Once we also put $m=h=0$ (leading to the
simplified elements $W(1,0,1)=-{w}\left( w+a \right)$ and
$Z(1,0,1,0)=i \left( a+w{a}^{2}- {w}^{3} \right)$), we may
immediately (and non-numerically) test the positivity of the matrix.

%\newpage
%********** Figure 1 zde
\begin{figure}[h]                     %instead of \begin{figure}[t]
\begin{center}                         %instead of \begin{center}
\epsfig{file=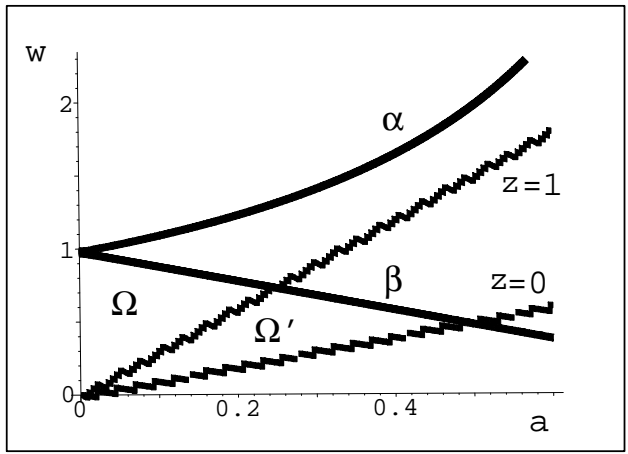,angle=270,width=0.6\textwidth}
\end{center}                         %instead of \end{center}
\vspace{-2mm} \caption{The thick-line upper boundaries
$\alpha=\alpha(a)$ and $\beta=\beta(a)$  of the respective
positivity domains of the respective eigenvalues $\theta_-^+$ and
$\theta_\pm^-$ of the eligible metric $\Theta\left [H^{(4)}(a,z)
\right ]_{(1,0,1,0)}$.
 \label{rm8jp}}
\end{figure}
%%\newpage

This test may proceed via an immediate evaluation of the four
eigenvalues $\theta_{\pm}^\pm$ of the candidate matrix $\Theta\left
[H^{(4)}(a,z) \right ]_{(1,0,1,0)}$ in the respective closed forms
 \ben
 \theta_+^\pm
 =1+  \frac{1}{2}\,\left (w-{a}^{2}w+{w}^{3}\right )
 \pm \frac{1}{2}\,\sqrt{\triangle^+}
 \label{prva}
 \een
and
 \ben
 \theta_-^\pm
 =1-  \frac{1}{2}\,\left (w-{a}^{2}w+{w}^{3}\right )
  \pm \frac{1}{2}\,\sqrt{\triangle^-}
 \label{druha}
 \een
with the common discriminant
 $$
 \triangle^\pm=
 {w}^{6}+ \left( 2-2\,{a}^{2} \right) {w}^{4}+ \left( \pm 8+4\,a \right) {w
}^{3}+ \left(5 \pm 8\,a+ 6\,{a}^{2}+{a}^{4} \right) {w}^{2}+
 $$
 $$
 +\left( 4\,a+4 \,{a}^{3} \right) w+4\,{a}^{2}\,.
 \ \ \ \ \ \ \ \ \ \ \ \ \ \
 \ \ \ \ \ \ \ \ \ \ \ \ \ \
 \ \ \ \ \ \ \ \ \ \ \ \ \ \
 $$
These formulae enable us to specify the parametric domain of the
necessary  positivity of the metric $\Theta\left [H^{(4)}(a,z)
\right ]_{(1,0,1,0)}$ by elementary means.

The results are sampled in Fig.~\ref{rm8jp} where we displayed the
part of the $(z,a)$ plane where $\theta_+^+>0$ so that just the two
eigenvalues may get non-positive there. The inspection of this
picture enables us to conclude that these eigenvalues remain
positive (so that the general pseudometric $\Theta \left[
H^{(4)}(a,z) \right ]_{(1,0,1,0)}$ becomes tractable as the {\em
positive definite} metric) in the triangular domain
$\Omega\bigcup\Omega'$ (assuming that $z>0$) or $\Omega$ (assuming
that $z>1$). Such a simplification follows from the fact that at the
two sample values of $z=0$ and of $z=1$, the auxiliary wiggly lines
just ``translate" the  $a-$dependence into $w-$dependence of the
change-of-sign boundary since the auxiliary functions $w=w(z,a)=
{3}^{z}{a}\,$ (which enter the above closed formulae as
abbreviations) are linear in $a$.

\section{The  $N=6$ model\label{nje6}}

For the not too large matrix dimensions $N=2K$ the technique of the
construction of the general pseudometrics $\Theta$ via the solution
of the linear algebraic Eq.~(\ref{dieudonne}) remains feasible and
straightforward. With the growth of $N$ the only difficulty emerges
in connection with the printed presentation of the resulting
multiparametric set of pseudometrics. For this reason it makes sense
to find a sufficiently representative set of a few key parameters.
For each such choice, moreover, it becomes necessary to determine a
boundary of the domain ${\cal D}$ of these key parameters, inside
which the pseudometric in question remains positive definite and,
hence, eligible as a metric in a certain ``optimal" physical Hilbert
space ${\cal H}^{(S)}$.

In our present setting the task is slightly simplified by the fact
that the most relevant values of our first, ``kinematical input"
parameter $a$ may be expected small and admitting, therefore, the
use of perturbation approximations. Secondly, we may take the
variability of the second, ``dynamical input" parameter $z$ for
granted. Nevertheless, as long as the choice of this exponent will
be mostly dictated by applications (controlling, e.g., the loss of
stability of certain most fragile levels), its choice proves not too
essential in our present methodical considerations. We shall often
consider the ``discrete square-well value" \cite{dsw} $z=0$ as a
sufficiently instructive and generic option.

\subsection{The  metric
at $z=0$}

 %\noindent
Even when we set our ``methodically redundant" exponent $z$ equal to
zero, we cannot parallel Eq.~(\ref{gene4}) and display the whole
six-parametric matrix $\Theta$ resulting from the computer-assisted
non-numerical solution of Eq.~(\ref{dieudonne}). For this reason we
further employed the simplification used in subsection \ref{n4a} and
demanded that all the elements of the main diagonal of our special
$\Theta$ are chosen equal to one. Although the resulting
pseudometric still did not fit in the printed page, the separate
matrix  elements do and remain sufficiently compact,
 $$
\Theta_{2,1}= \Theta_{6,5}=m+ia\,,\ \ \ \
\Theta_{3,2}=\Theta_{5,4}=4\,m{a}^{2}+m+d+2\,ia \,,
 $$
 $$\ \ \
\Theta_{4,3}= d+4\,m{a}^{2}+m+r+3\,ia\,,
 $$
 $$
\Theta_{3,1}= \Theta_{6,4}=-2\,{a}^{2}+2\,ima\,,\ \ \ \
\Theta_{4,2}=\Theta_{5,3}=-6\,{a}^{2}-i \left(
-8\,m{a}^{3}-4\,ma-2\,da \right)\,,
 $$
 $$
\Theta_{4,1}= \Theta_{6,3}=d-i \left( 4\,{a}^{3}-a \right) \,,\ \ \
\ \Theta_{5,2}=r+d-i \left( -2\,a+4\,{a}^{3} \right) \,,
 $$
 $$
\Theta_{5,1}= \Theta_{6,2}=-4\,{a}^{2}-i \left(
-2\,ma-8\,m{a}^{3}-2\,da \right)  \,,\ \ \ \ \Theta_{6,1}=r+ia\,.
 $$
The complex conjugates of these elements form the upper-triangle
part of $\Theta=\Theta^\dagger$ of course.

%\newpage
%********** Figure 1 zde
\begin{figure}[h]                     %instead of \begin{figure}[t]
\begin{center}                         %instead of \begin{center}
\epsfig{file=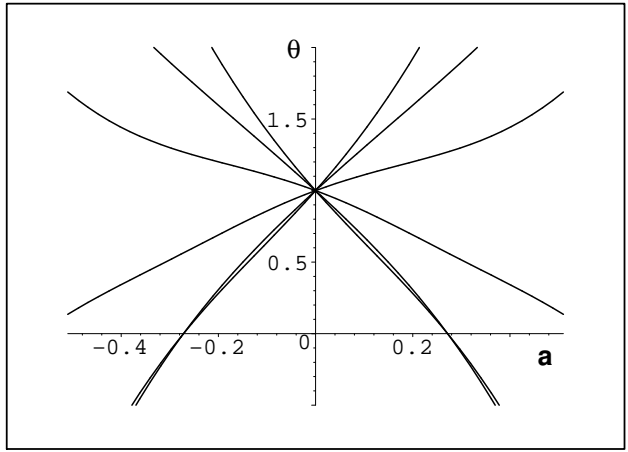,angle=270,width=0.6\textwidth}
\end{center}                         %instead of \end{center}
\vspace{-2mm} \caption{The spectrum of the simplified metric
(\ref{hvezda}).
 \label{drm0}}
\end{figure}
%%\newpage

In a continuation of the search of parallels with the results of
subsection \ref{n4a} we may further set $m=d=r=0$ and obtain
$\Theta=\Theta^{(6)}_0(a)$ as the sufficiently compact matrix
 \be
 \left[ \begin {array}{cccccc}
 1&-ia&-2\,{a}^{2}&i \left( 4{a}^{3}-a
 \right) &-4\,{a}^{2}&-ia
 \\\noalign{\medskip}ia&1&-2\,ia&-6\,{a}^{2}&i
 \left( 4{a}^{3}-2a \right) &-4\,{a}^{2}
 \\\noalign{\medskip}-2\,{
 a}^{2}&2\,ia&1&-3\,ia&-6\,{a}^{2}&i \left( 4{a}^{3}-a \right)
 \\\noalign{\medskip}i \left( a-4{a}^{3} \right) &-6\,{a}^{2}&3\,ia&
 1&-2\,ia&-2\,{a}^{2}
 \\\noalign{\medskip}-4\,{a}^{2}&i \left(
 2a-4 {a}^{3} \right)
 &-6\,{a}^{2}&2\,ia&1&-ia
 \\\noalign{\medskip}ia&-4\,{ a}^{2}&i
 \left( a-4{a}^{3} \right) &-2\,{a}^{2}&ia&1\end {array}
 \right].
 \label{hvezda}
 \ee
The graph of its eigenvalues is displayed in Fig.~\ref{drm0},
showing that such a matrix may serve as a metric in Hilbert space
iff $a \in (-\beta^{(6)},\beta^{(6)})$ with $\beta^{(6)}\approx
0.2718445$.

%\newpage
%********** Figure 1 zde
\begin{figure}[h]                     %instead of \begin{figure}[t]
\begin{center}                         %instead of \begin{center}
\epsfig{file=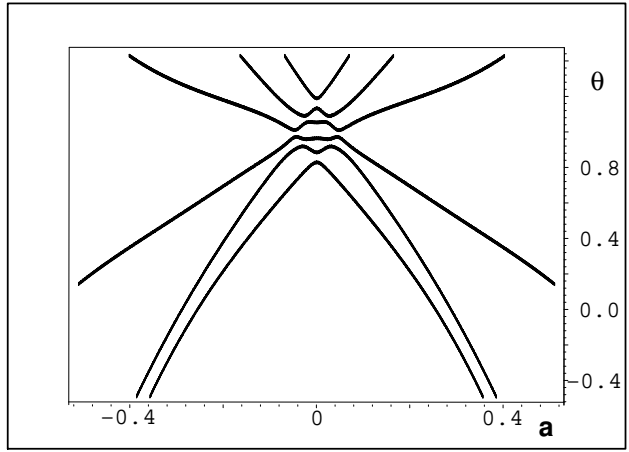,angle=270,width=0.6\textwidth}
\end{center}                         %instead of \end{center}
\vspace{-2mm} \caption{The spectrum of the perturbed metric
(\ref{hvezda}) with $m = 1/10$.
%$\Theta\left [H^{(6)}(a,z) \right ]_{(1,1/10.0.0)}$.
 \label{drm1l10}}
\end{figure}
%%\newpage

The inspection of Fig.~\ref{drm0} reveals that in the domain of very
small $a$s the $a-$dependence of the eigenvalues $\theta_j(a)$ of
the metric $\Theta_0^{(6)}(a)>0$ may be very well approximated by
the linear functions, $\theta_j(a)\approx 1+h_j\,a$. The
determination of the coefficients $h_j$ is still easy since we may
Taylor-expand the metric
 \be
  \Theta(a)=I + \left[ \begin {array}{cccccc} 0&-ia&0&-ia&0&-ia
  \\\noalign{\medskip}ia
 &0&-2\,ia&0&-2\,ia&0
 \\\noalign{\medskip}0&2\,ia&0&-3\,ia&0&-ia
 \\\noalign{\medskip}ia&0&3\,ia&0&-2\,ia&0
 \\\noalign{\medskip}0&2\,ia&0
 &2\,ia&0&-ia
 \\\noalign{\medskip}ia&0&ia&0&ia&0
 \end {array}
 \right]+{\cal O}\left (a^2\right ) \,
 \label{lihvezda}
 \ee
and arrive at the virtually trivial leading-order secular equation
 $$
 h_j^{6}-26\,h_j^{4}+181\,h_j^{2}-225=
(h_j^3-2\,h_j^2-11\,h_j+15)\,(h_j^3+2\,h_j^2-11\,h_j-15)=0
 $$
and roots $h_{1,6}= \mp 3.102940862$, $h_{2,5}=\pm 1.256913500$,
$h_{3,4}=\pm 3.846027361$.

The practical use of such a result may be twofold:

\begin{itemize}

\item
we may assure ourselves that the decrease of the lowest eigenvalue
$\theta_{j_{\rm min}}(a)\approx 1-3.846\,a$ {\em guarantees} that
the maximal admissible value of $a$ is $\beta \approx 1/3.846027361
\approx 0.2600$ plus/minus ${\cal O}(a^2)$ corrections; this
estimate seems fully consistent with the above-derived exact value
of $\beta^{(6)} \approx 0.2718445$;

\item
alternatively, we may restrict ourselves to a much smaller
subinterval of $a$ in which the exact minimal eigenvalue
$\theta_{j_{\rm min}}(a)$ remains safely positive, say, on the
grounds of a variational estimate of corrections (keeping also in
mind that $\max a^2 \approx 0.06$).

\end{itemize}

 \noindent
The formalism of perturbation expansions may  be recalled and used
to reflect the influence of further parameters. The importance of
the possible changes of the spectrum of the metric is sampled in
Fig.~\ref{drm1l10} where the differences from Fig.~\ref{drm0} are
all caused by the choice of $m=1/10$. One witnesses the complete
removal of the full zero-order degeneracy of the ``unperturbed"
$\theta_j=1+ {\cal O}(a)$. Just a partial removal of the degeneracy
may result from the alternative choices, e.g., of $r\neq 0$ which
just splits the six-times degenerate $a=0$ eigenvalue $\theta_j$ in
two three-times degenerate descendants.

\section{Extrapolations and models with  $N\geq 8$\label{nje8}}

At any dimension $N=2K$ the dynamical contents of our toy models is
controlled by the Hamiltonian (which varies with the ``kinematical"
parameter $a\in (-\alpha^{(N)},\alpha^{(N)})$, ``dynamical"
parameter $z\in (-\infty,\infty)$ and real spectral locus
$\{\varepsilon_j(a,z)\}$) {\em and} by the Hermitian,
positive-definite metric $\Theta^{(N)}$ specified by an $N-$plet of
parameters $\{k, m, \ldots\} \in {\cal D}$.

Naturally, practical implementations of such a recipe will require
also a determination of the whole positivity domain ${\cal D}$, or
of its suitable non-empty subdomain at least. This task will
certainly be facilitated by the smallness of $a$ in practice. In our
constructions performed at $N \leq 6$ we saw, moreover, that many of
the closed small$-a$ formulae might be potentially extrapolated to
an arbitrary dimension $N=2K$.

\subsection{The $N\geq 8$ metrics \label{menje8}}

One of the most promising keys to the extension of at least some of
our above-listed low-dimensional results to all $N$ has been found
in the compact form of the matrix elements of the pseudometrics at
$N\leq 6$. Indeed, we succeeded in transmuting these small$-K$
elements into larger$-K$ ansatzs and found out that such a method of
construction appeared very efficient and quick.

One of the main shortcomings of such a recipe lies in the enormous
growth of the size of the formulae caused, mainly, by the linear
growth of the number of free parameters with the increasing
dimension $N=2K$. This means that in the computer-assisted
environment we may still deduce the form of the $N-$parametric
pseudometrics $\Theta^{(N)}$ (from Eq.~(\ref{dieudonne})) but the
resulting extrapolations of the above-displayed compact $K=1$, $K=2$
and $K=3$ formulae cease to be compact. Thus, for presentation
purposes, the majority of the available $N \geq 8$ results still has
to be compactified and transformed, typically, into a graph or
numerical table.

During the computer-assisted $N \geq 8$ constructions themselves,
the most difficult obstacle has been found, as we already indicated,
in the necessary specification of the boundaries of a non-empty
metric-positivity (sub)domain ${\cal D}^{(N)}$. Fortunately, the
very natural assumption of smallness of $a$ almost trivialized the
problem at $N\leq 6$ and proved also helpful at the higher
dimensions. The point was that the advantage of the practical
negligibility of $a^2$ shortened the solution of
Eq.~(\ref{dieudonne}), implying the -- unexpected -- easiness of the
extrapolation of the matrix elements of pseudometrics to all $N$.

As we have already noted, the complete, multiparametric matrices
$\Theta$ can hardly be displayed, in print, even at $K=3$. For this
reason, let us restrict our attention to certain special subsets of
metrics and present the theory ``via examples". Firstly, let us skip
the questions of energies (discussed, at length, in the preceding
sections after all) and study, from now on, just the simplest
possible toy-model dynamics with $z=0$.

Secondly, let us circumvent the (difficult) problem of the
determination of the exact ``exceptional point" boundaries
$\partial{\cal D}^{(N)}$ (redirecting the interested readers, say,
to our dedicated study \cite{maximal}) and let us select just a
single representative point (i.e., multiindex $\mu:=(k,m,\ldots)\to
(k_0,m_0,\ldots):=\mu_0$) which lies safely inside the hypothetical
parametric domain ${\cal D}^{(N)}$.

Thirdly, the guarantee of the latter requirement $\mu_0\in {\cal
D}^{(N)}$ will be made easy via the extension of our previous $N\leq
6$ experience to all $N$ and by the selection of the main-diagonal
elements of the candidate for the metric equal strictly to one,
$\Theta_{n,n}=1$, $n=1,2,\ldots,N$ (i.e., $k_0=1, \ldots$), with all
of the other parameters staying ``sufficiently small"  (i.e., $|m_0|
\ll 1, \ldots$).

\subsection{The $N \geq 8$ metrics at $z=0$ and at small $a$ \label{umenje8}}

In a continuation of our simplified presentation of the
extrapolation ideas let us now select the dimension $N=8$ and set
all of the ``small" parameters in a general pseudometric $\Theta$
strictly equal to zero (i.e., $m_0=0, \ldots$). Immediately, the
explicit solution of Eq.~(\ref{dieudonne}) will generate the
pseudometric $\Theta^{(8)}_0$ with matrix elements which become
predictable, by extrapolation, from their above-displayed $N=6$
predecessors.

On this basis we verified, numerically, that also the $a-$dependence
of the spectrum $\{\theta_j\}$ of the new pseudometric
$\Theta^{(8)}_0(a)$ remains very similar to the one sampled in
Fig.~\ref{drm0} above. We deduced, extrapolated and also re-verified
at $N=10$ that at any $N=2K$, one may expect and conjecture to see
the $K-$plet of curves $\theta_j(a)$ which are moving quickly up
with the growth of $|a|$, complemented by another $K-$plet of curves
$\theta_j(a)$ which are moving quickly down with the growth of
$|a|$.

At this moment one can recollect the $N=6$ discussion of
Fig.~\ref{drm0}  and formulate the following two questions
concerning the determination of the pseudometric-positivity interval
of $a \in (-\beta^{(N)},\beta^{(N)})$ at the general dimension:

\begin{itemize}

\item
do the left and right intersections $\pm \beta^{(N)}\in
\partial{\cal D}^{(N)}$ of the lowest eigenvalue curve
$\theta_{j_{\rm min}}(a)$ with the $a-$axis leave a non-empty and/or
sufficiently large space for the variability of $a$ at large $N=2K$?

\item
would the  linear approximation $\theta_{j_{\rm min}}(a)=
1+h_{j_{\rm min}}\,a\ $ of the minimal eigenvalues provide a
sufficiently reliable estimate of the exact exception-point values
of $\beta^{(N)}$?

\end{itemize}

Whenever both of the answers happen to be positive, we may proceed
further,  accept the above-introduced perturbation-approximation
philosophy and linearize our pseudometric. Thus, we returned to
Eq.~(\ref{dieudonne}), constructed the linearized matrix
$\Theta^{(8)}_0$ and obtained it in the following, very regular
sparse-matrix form
 $$
\left[ \begin {array}{cccccccc}
1&-ia&0&-ia&0&-ia&0&-ia\\\noalign{\medskip}ia&1&-2\,ia&0&-2\,ia&0&-2\,ia&0
\\\noalign{\medskip}0&2\,ia&1&-3\,ia&0&-3\,ia&0&-ia
\\\noalign{\medskip}ia&0&3\,ia&1&-4\,ia&0&-2\,ia&0\\\noalign{\medskip}0
&2\,ia&0&4\,ia&1&-3\,ia&0&-ia\\\noalign{\medskip}ia&0&3\,ia&0&3\,ia&1&
-2\,ia&0\\\noalign{\medskip}0&2\,ia&0&2\,ia&0&2\,ia&1&-ia
\\\noalign{\medskip}ia&0&ia&0&ia&0&ia&1\end {array} \right]\,.
 $$
We see that the extrapolation of this leading-order form of the
pseudometric to any $N=2K$ is truly trivial. Also the construction
of its linear-approximation eigenvalues $\theta_{j_{\rm }}(a)=
1+h_{j_{\rm }}\,a\ $ remains feasible at any dimension. For
illustration let us just select $N=8$ and display the related
leading-order secular equation for the coefficients,
 $$
h_j^{8}-70\,h_j^{6}+1487\,h_j^{4}-9139\,h_j^{2}+11025=0\,.
 $$
It may again be factorized yielding the two parallel rules
 $$
  h_j^{4}\mp 2\,h_j^{3}-33\,h_j^{2}\pm 47\,h_j+105 =0
 $$
and roots 1.259204635, 2.752948888, 5.256297172 and 5.762552919 of
both signs which determine the leading-order eight-line-crossing
small$-a$ part of the $N=8$ analogue of the $N=6$ spectra of
Fig.~\ref{drm0}.

We may add that the linear-extrapolation prediction $1/5.76255
\approx 0.1735$ of the exceptional-point value as obtained from the
maximal root again compares very well with the exact numerical value
of $\beta^{(8)} \approx 0.1683983$. Thus, the linear approximation
leads to the error $\sim 0.005$ which happens to be much smaller
than the rough estimate $\sim \left [\beta^{(8)}\right ]^2 \approx
0.03$ of the typical magnitude of the second order correction.

\subsection{Discussion: the $N \geq 8$ discrete approximants and an efficient
spectral design\label{discussion}}

Our present proposal of working with discrete, matrix quantum models
of finite dimension has been initiated by certain formal
difficulties encountered, in the literature, during the attempted
assignments of a nontrivial Hilbert-space metric to a given,
non-Hermitian differential Hamiltonian operator possessing the
strictly real spectrum. Typically, these difficulties are being
circumvented via additional assumptions (cf. our comments above,
especially in section \ref{expla}).

During the very initial step of our analysis (viz., during the
replacement of Eq.~(\ref{SEloc}) by Eq.~(\ref{SEdis}) using a finite
number of grid points $N=2K$) an elementary parameter $a$ emerged
and specified a fixed length in our models. In comparison with the
differential-operator predecessors of our models, a new quality
emerged since the spectra of energies became ``fragile", i.e.,
complex beyond a certain critical kinematics-related size-parameter
$a_{critical} =\alpha^{(N)}>0$.

From the point of view of the variability of dynamics our initial
choice of the one-parametric discrete versions of the very special
potential of the imaginary cubic oscillator (as made in section
\ref{energies}) did not prove too satisfactory. Fortunately, we
managed to shift the role of the most unstable state to optional
excitations by means of a re-scaling of the potential based on an
introduction of another, ``dynamical" exponent-parameter $z \in
\mathbb{R}$.

Undoubtedly, the latter trick made the structure (and, in
particular, the ``topological menu") of the energy levels
``universal" in the sense illustrated, at $N=8$, in
Table~\ref{roxp4}. At the same time, the mechanism of the changes of
the topological structure of the spectral loci $\varepsilon_j(a)$
remained transparent and tractable, schematically, as an up and down
``motion" of the two (deformed) circles. In this context the
information compressed in Table~\ref{roxp4} was complemented,
graphically, by a series of illustrative pictures. All of these
observations will find their strict analogues at any integer $N =
2K$.

Another merit of our toy-model simulations of dynamics may be seen
in the related feasibility of the constructions and in the
extrapolation-friendliness of the multiparametric matrices of the
pseudometrics. This implies, certainly, the rarely encountered and
equally rarely employed freedom of the control of dynamics by the
metric and of the related, rarely emphasized \cite{Geyer}
theoretical possibility of the {\em ad hoc} modifications of the
classes of the additional observables $\Lambda_1, \Lambda_2,
\ldots$.

An important further merit of our present models has been found in
the feasibility of the computer-assisted generation and {\em
extrapolations} of important formulae (as well as of their graphical
pendants and topological interpretations) {\em to all $N$}. As an
immediate consequence one must appreciate, among others, the
facilitated estimates of the ranges of the positivity of the
pseudometrics, or the facilitation of the applicability of the
linear-algebraic and perturbation-expansion techniques.

Many of these ideas may find further applications. At the same time,
we would expect that in the spirit, say, of Refs.~\cite{fund,fundb},
the next-step developments of the subject should be aimed at the
simulations of dynamics which replace the local potentials $V(x)$ by
some slightly non-local generalizations. Also in this respect, the
present technique of discretization may be expected to lower many
purely technical obstacles.

\section{Summary\label{summary}}

Our first tests of the idea of discretization proved disappointing.
We observed that the growing-dimension series (\ref{dva}) of the
simple-minded one-parametric descendants of the popular differential
imaginary cubic oscillator do not offer a sufficiently rich
variability of the coupling-dependence of the energy spectra.
Fortunately, the merely slightly more sophisticated and re-scaled
two-parametric choice (\ref{zzz}) of the model has been shown to
offer a flexibility of the spectral loci which covers a broad menu
of alternative mechanisms of the phenomenologically interesting
tunability.

In our two-parametric model the breakdown of stability of the system
was shown to be caused by the spontaneous, dynamically controlled
complexification which could be designed as destroying the reality
and stability of any pre-selected pair of neighboring bound states.

The price to be paid for such a highly welcome universality of the
model lies in the necessity of a rather complicated construction of
the physical {\em ad hoc} metrics $\Theta$. In this setting we took
the advantage of the efficiency of the direct solution of the
Dieudonn\'{e}'s Eq.~(\ref{dieudonne}) as reported in
Ref.~\cite{fund}. We discussed the related idea of reduction of the
ambiguity of the metric via the requirement of its maximal
computational friendliness, i.e., of its sparse-matrix structure. We
tested this possibility and arrived at encouraging results here.

One of the main formal advantages of our present class of models may
be seen in the possibility of its detailed study at the smallest
dimensions  followed by the formulation of ansatzs an by their tests
and successful trial-and-error extrapolations to {\em arbitrary}
dimensions. A particularly efficient application of such a strategy
has been found in the context of perturbation-series constructions
where we made use of the fact that due to its
Runge-Kutta-approximation origin, the parameter $a$ may be truly
considered very small in practice.

Our analysis also confirmed expectations that for the real exponents
$z$, the metric operators can never be diagonal or banded matrices.
This seems to be a characteristic consequence of the choice of a
local form of the interaction for which the inner products remain
``long-ranged" in the sense explained, in the context of scattering
theory, by Jones \cite{Jonesdva}. Thus, in accord with our
commentary in~\cite{scatt} we believe that the requirement of the
unitarity of the scattering implies the necessity of introduction of
at least small non-localities in the interaction. In opposite
direction we expect that our present complex local-like Hamiltonians
will preserve the analogy with their differential-operator-like
$N\gg 1$ limits. In particular, we are persuaded (and would like to
conjecture) that these models of dynamics will never admit the
existence of a unitary and causal version of the scattering, not
even in the ``short-range" dynamical regime with very negative
exponents $z\ll 0$.

\section*{Acknowledgements}

Work supported by the GA\v{C}R grant Nr. P203/11/1433, by the
M\v{S}MT ``Doppler Institute" project Nr. LC06002 and by the
Institutional Research Plan AV0Z10480505.

%\newpage

\end{document}